\documentclass[twocolumn,showpacs,assymb,amsmath]{revtex4}
\usepackage{mathrsfs}
\usepackage{txfonts}
\usepackage{graphicx}
\usepackage{amssymb}
\usepackage{amsmath}
\usepackage{color}

\usepackage{graphicx}
\usepackage{color}

\begin{document}

\title{Effects of system-bath coupling on Photosynthetic heat engine: A polaron master equation approach}

\author{M. Qin$^{1,2}$, H. Z Shen$^{1}$, X. L. Zhao$^{1,2}$, and X. X. Yi$^1$
\footnote{Corresponding address: yixx@nenu.edu.cn}}
\affiliation{$^1$Center for Quantum Sciences and School of Physics, Northeast
Normal University, Changchun 130024, China\\
$^2$School of Physics and Optoelectronic Technology\\
Dalian University of Technology, Dalian 116024, China}

\begin{abstract}
In photosynthesis, the fundamental principles responsible for the near-unit quantum efficiency of the
conversion of solar to chemical energy remains unknown. Under natural conditions, the formation of stable
charge separation states in bacteria and plant reaction centers is strongly affected by the coupling of
electronic degrees of freedom to a wide range of vibrational motions. These inspire and motivate us to
explore the effects of environment on the operation of such complexes. In this paper, we apply the
polaron master equation, which offers the possibilities to interpolate between weak and strong
system-bath coupling, to study how system-bath couplings affect charge transfer processes in Photosystem
II reaction center (PS{\rm II} RC) inspired quantum heat engine (QHE) model in a wide parameter range. The effects of bath correlation and temperature, together
with the combined effects of these factors are also discussed in details. The results show a variety of
dynamical behaviours. We interpret these results in terms of noise-assisted transport effect and
dynamical localization which correspond to two mechanisms underpinning the transfer process in
photosynthetic complexes: One is resonance energy transfer and the other is dynamical localization effect captured by the polaron master equation. The effects of system-bath coupling and bath correlation are incorporated in
the effective system-bath coupling strength determining whether  noise-assisted transport effect or
dynamical localization dominates the dynamics and temperature modulates the balance of the two mechanisms. Furthermore, these two mechanisms can be attributed to one physical
origin: bath-induced fluctuations. The two mechanisms is manifestations of dual role played by bath-induced
fluctuations within respective parameter range. In addition, we find that the effective voltage of QHE exhibits superior robustness with respect to the bath noise as long as the system-coupling strength is not too large.
\end{abstract}

 \maketitle
\section{Introduction}
In the photosynthesis process of plants and bacteria, the sun's energy is captured and stored by
a series of events that convert the pure energy of light into the biochemical energy
needed to power life, providing all our food and most of our energy resources
\cite{Amerongen2000,Blankenship2002}. After
absorption of a photon, an excited state on a pigment molecule is created. The excitation is transferred
efficiently through antenna system until it arrives at a reaction center (RC) where charge separation and conversion of pure energy of excited states to  chemical changes take place
\cite{Blankenship2011332,Wientjes2013117}. Numerous studies focus on the precise mechanisms underlying the
high efficiency transport
\cite{Hayes201012,Scholes20113,Lambert20139,Omar2014}. Recent experimental demonstrations of oscillatory electronic dynamics in photosynthetic system have raised the
quantum coherent dynamics may be relevant in photosynthetic energy transfer of living organisms in conditions that are often defined as hot and wet
\cite{Brixner2005434,Engel2007446,Calhoun2009113,Collini2010463,
Panitchayangkoon2010107,Harel2012109,Romero201410,Fuller20146}. This causes much debate whether quantum coherence
promotes the efficiency and the role of environment on energy transfer are also extensively discussed \cite{Olaya200878,Plenio200810,Mohseni2008129,Rebentrost200911,Caruso2009131,Ishizaki2009106,
Chin201012,Scholes20101,Lloyd201012,Strumpfer20123,Dong20121,Rey20134,Chin20139,Huelga201354}.
Therefore, understanding the underlying mechanism of such highly efficient excitation energy transduction in natural photosynthetic system can assist us in improving the design of promising artificial structures
for quantum transport and optimized light-harvesting devices \cite{Scully2010104,Scully2011108,Zhang201517,Ajisaka20155}.

Exposed to sunlight, RC complexes operate as Nature's solar cells with very high yield for light-to-charge conversion. Charge separation in RCs has been a question of recent studies. Dorfman et al have introduced a
promising approach in which the photosynthetic reaction center is viewed as a quantum heat engine (QHE),
and showed that attributed to noise-induced quantum coherence, the photocurrent of the photocell based on Photosystem {\rm II} reaction center (PS{\rm II} RC) can be increased by $27\% $ compared to an equivalent classical photocell \cite{Dorfman2013110}. Ref. \cite{Stones2016} investigates the effects of structured environment on electron transfer in PS{\rm II} RC-based photocell devices placed between two electrodes. In experiment, steady-state and multi-dimensional optical spectroscopy
have revealed that the process of excitation energy transfer and conversion to stable charge separated states is strongly affected by the interaction between excitonic or electronic degrees of freedom to a wide
range of vibrational modes of surrounding bath \cite{Renger2011104,Novoderezhkin200589,Novoderezhkin2004108}. In spite of these theoretical and experimental efforts, the fundamental principles responsible for charge transfer in PS{\rm II} reaction center are still indistinct and under scrutiny. These motivate us to consider the effect of system-bath coupling on charge transfer processes in PS{\rm II} RC.

Traditionally, transfer processes in open system has been described by F\"{o}rster-Dexter theory \cite{Forster195927,Dexter195221} if the electronic coupling between chromophores is very week compared with their interaction with bath. When the electronic coupling is strong and the system-bath coupling is
weak, it is necessary to consider relaxation between delocalized exciton states. In this limit, excitation energy transfer (EET) dynamics are
described by the coupled Redfield equations \cite{Redfield19651,Redfield19571}. However, for the intermediate coupling regime where the energy scales of electronic coupling and exciton-bath interaction are comparable, both F\"{o}rster-Dexter and Redfield theory become invalid since a proper perturbative term in theoretical treatment does not exist. In spit of the invalidity for the intermediate coupling case,
these two second-order perturbative theories also have a problem of being not precise enough according to recent spectroscopic experiments on photosynthetic complexes \cite{Engel2007446,Collini2010463,Panitchayangkoon2010107}. This calls for non-perturbative techniques to
obtain numerically exact dynamics, for example, the quasi-adiabatic propagator path integral (QUAPI) \cite{Makri199500,Makri199511}, the hierachy equations of motion (HEOM) \cite{Tanimura200675,Ishizaki200574} and the multiconfiguration time-dependent Hartree approach \cite{Meyer1990165,Beck2000324,Thoss2001115}. Nevertheless, these methods are computationally sophisticated and not trivial to implement especially for large system size or multi-excitation case. Thus, a computationally economical and an appropriate qualitative and quantitative account of dynamics for the intermediate case is in urgent need.

Recently, a polaron transformed second-order master equation has been developed to treat coherent energy transfer in molecular system \cite{Jang2008129,Jang2009131,Jang2011135,Nazir2009103,McCutcheon201183,Kolli2011135,Silbey198480,
McCutcheon2011135,McCutcheon201184,Pollock201315,Grover197154,Silbey198072,Chang2012137,Kolli2012137,
Zimanyi2012370,Lee2012136,Xu201618}. Despite perturbative, this approach allows for a consistent exploration of the
intermediate regime in which many multichromophoric systems operate, serving as a bridge between the Redfield and F\"{o}rster-Dexter theories. In this formulism, the system-plus-phonon bath Hamiltonian is transformed into the polaron frame, in which the system Hamiltonian is dressed by a phonon and electronic couplings are renormalized and fluctuate due to coupling with phonon bath. Then the transformed system-bath interaction term can be treated as a perturbation and the master equation can be obtained using standard projection operator techniques. This approach combines the excitation with its surrounding bath as an entity instead of considering the exciton and bath separately and has been reliably applied to describe EET in light-harvesting complexes in the intermediate coupling regime as a quantitative method. In this paper, we apply this approach to shed light on the question how system-bath couplings affect charge transfer processes in PS {\rm II} RCs in a wide parameter range.

The remainder of the paper is organized as follows: In Sec. {\rm
II}, we introduce a model of PS{\rm II} RC-based quantum heat engine
to describe the charge separation. Then we review the formulism of polaron transformation and give polaron master equation for our model. The concepts of effective voltage and power generated applied for assesing the performance of our QHE system is also introduced in this
section. In Sec. {\rm III}, we first elucidate the temperature and system-bath coupling strength dependence of the effective coupling and its effect on the equilibrium structure of the exciton-ICTS system, and then  focus on the separate effects of system-bath coupling $\gamma$, cross-correlation coefficient $c$ and temperature $T$, respectively. Furthermore we discuss the combined effects of these parameters on the current and power generated by QHE system in details. We utilize Franck-Condon factor and the concept of effective system-bath coupling including the effects of both individual system-bath coupling and bath correlation to explain the various behaviours of QHE performance. Two mechanisms dominating the transfer dynamics are concluded. In addition, we explore the effective voltage of QHE model subjected to bath noise. Sec. {\rm IV} is
devoted to concluding remarks. We leave the detailed derivation of polaron master equation
in the appendix.
\section{Theory}
\subsection{Model system}
\begin{figure*}
\centering
\includegraphics[scale=0.2]{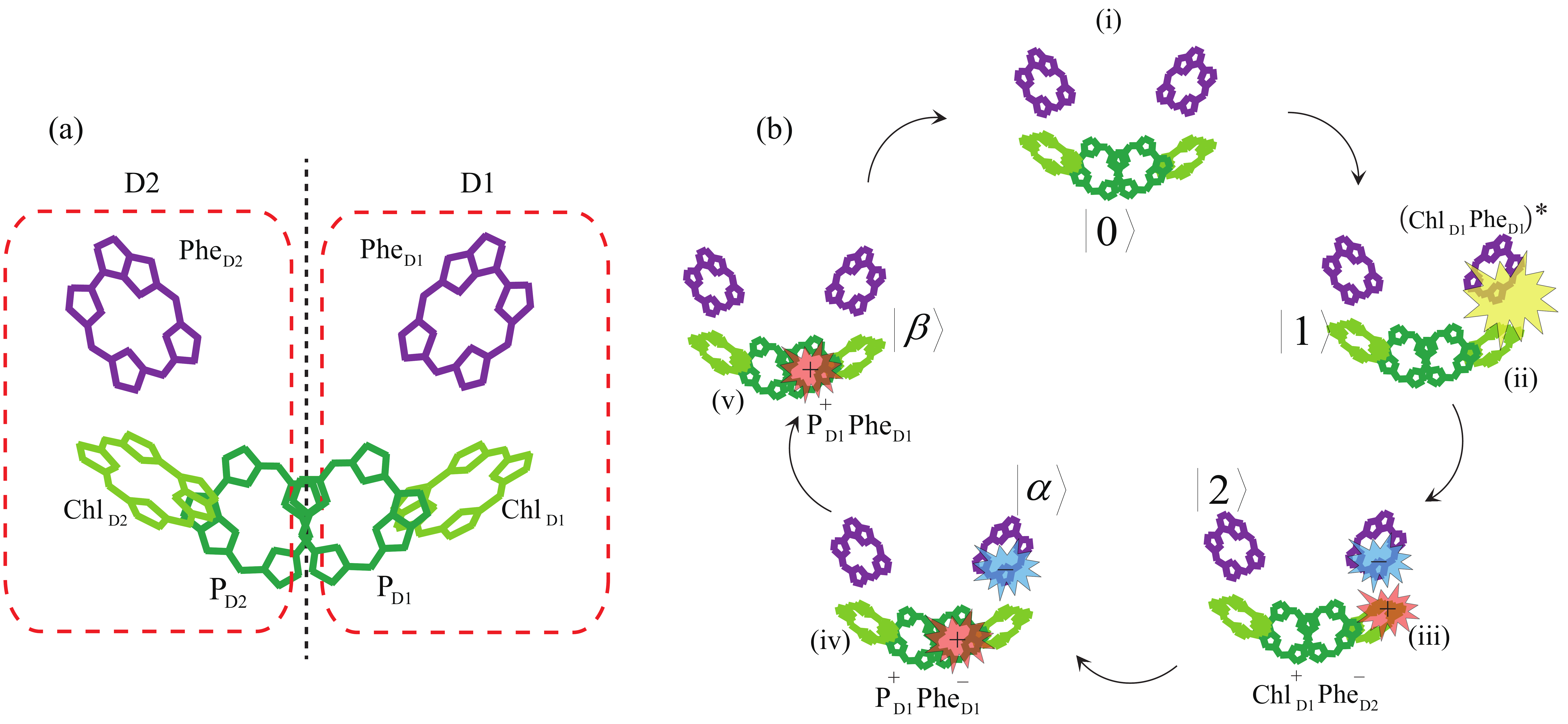}
\caption{(Color online)
(a) Arrangement of six core-pigments in the
PS{\rm II} RC. It consists of a special pair
${{\rm{P}}_{{\rm{D1}}}}$, ${{\rm{P}}_{{\rm{D2}}}}$, two
accessory chlorophylls ${\rm{Ch}}{{\rm{l}}_{{\rm{D1}}}}$,
${\rm{Ch}}{{\rm{l}}_{{\rm{D2}}}}$, and two pheophytins
${\rm{Ph}}{{\rm{e}}_{{\rm{D1}}}}$,
${\rm{Ph}}{{\rm{e}}_{{\rm{D2}}}}$.
(b) Charge transfer process in our QHE model.
(i) The lowest energy configuration with all six pigments in neutral ground state after an electron has been replenished at the special pair.
(ii) Excited state ${\left( {{\rm{Ch}}{{\rm{l}}_{D1}}{\rm{Ph}}{{\rm{e}}_{D1}}} \right)^ * }$ after absoption of a photon.
(iii) Primary charge transfer state ${\rm{Chl}}_{D1}^ + {\rm{Phe}}_{D1}^ - $ after ${{\rm{Ch}}{{\rm{l}}_{D1}}}$ as the electron donor rapidly loses an electron to the nearby electron acceptor molecule ${{\rm{Ph}}{{\rm{e}}_{D1}}}$.
(iv) Secondary charge transfer state ${\rm{P}}_{D{\rm{1}}}^ + {\rm{Phe}}_{D1}^ - $ after the positive and negative charges are spatially separated.
(v) Positively charged state ${\rm{P}}_{D{\rm{1}}}^ + {\rm{Phe}}_{D1}^{}$ after an electron has been released from the system to perform work.}
\label{structureme:}
\end{figure*}

We first illustrate the structure of PS{\rm II} RC complex in Fig.~\ref{structureme:} (a).
PS{\rm II} RC contains four chlorophylls (special pair ${{\rm{P}}_{D1}}$ and ${{\rm{P}}_{D2}}$ and accessory ${{\rm{Ch}}{{\rm{l}}_{D1}}}$ and ${{\rm{Ch}}{{\rm{l}}_{D2}}}$) and two pheophytins (${\rm{Ph}}{{\rm{e}}_{D1}}$ and ${\rm{Ph}}{{\rm{e}}_{D1}}$) arranged in two branches (${{\rm{D}}_1}$ and ${{\rm{D}}_2}$). Only ${{\rm{D}}_1}$ branch takes an active part in the electron transfer process. Although two different excited sates ${\left( {{{\rm{P}}_{D1}}{{\rm{P}}_{D2}}{\rm{Ch}}{{\rm{l}}_{D1}}} \right)^ * }$ and ${\left( {{\rm{Ch}}{{\rm{l}}_{D1}}{\rm{Ph}}{{\rm{e}}_{D1}}} \right)^ * }$ are discovered to initiate charge separation via ${{\rm{Ch}}{{\rm{l}}_{D1}}}$ and ${{\rm{P}}_{D1}}$ pathways respectively, we focus on the ${{\rm{Ch}}{{\rm{l}}_{D1}}}$ pathway, since experimental spectroscopy and theoretical researches reveals that charge transfer is mostly initiated from ${\left( {{\rm{Ch}}{{\rm{l}}_{D1}}{\rm{Ph}}{{\rm{e}}_{D1}}} \right)^ * }$ \cite{Novoderezhkin201112}. As shown in Fig.~\ref{structureme:} (b),  ${\rm{Ch}}{{\rm{l}}_{D1}}$
rapidly loses an electron to ${\rm{Ph}}{{\rm{e}}_{D1}}$, generating an initial charge transfer state (ICTS) $\left| {{\rm{Chl}}_{D1}^ + {\rm{Phe}}_{D1}^ - } \right\rangle $. Then the positive and negative charges are spatially separated to produce a secondary  charge transfer state $\left| {{\rm{P}}_{D1}^ + {\rm{Phe}}_{D1}^ - } \right\rangle $ which leads to energy stabilization. Then the electron is released from the system to drive a chain of chemical reactions including the reduction of NADP to
NADPH, the synthesis of ATP and the oxidized part of the RC splits
water, releasing molecular oxygen, and the system is positively charged denoted by ${\rm{P}}_{D{\rm{1}}}^ + {\rm{Phe}}_{D1}^{}$. At last, the system captures an electron from
their surroundings to complete the cycle and returns to the
ground state in which none of the six pigment molecules are excited.

\begin{figure*}
\centering
\includegraphics[scale=0.3]{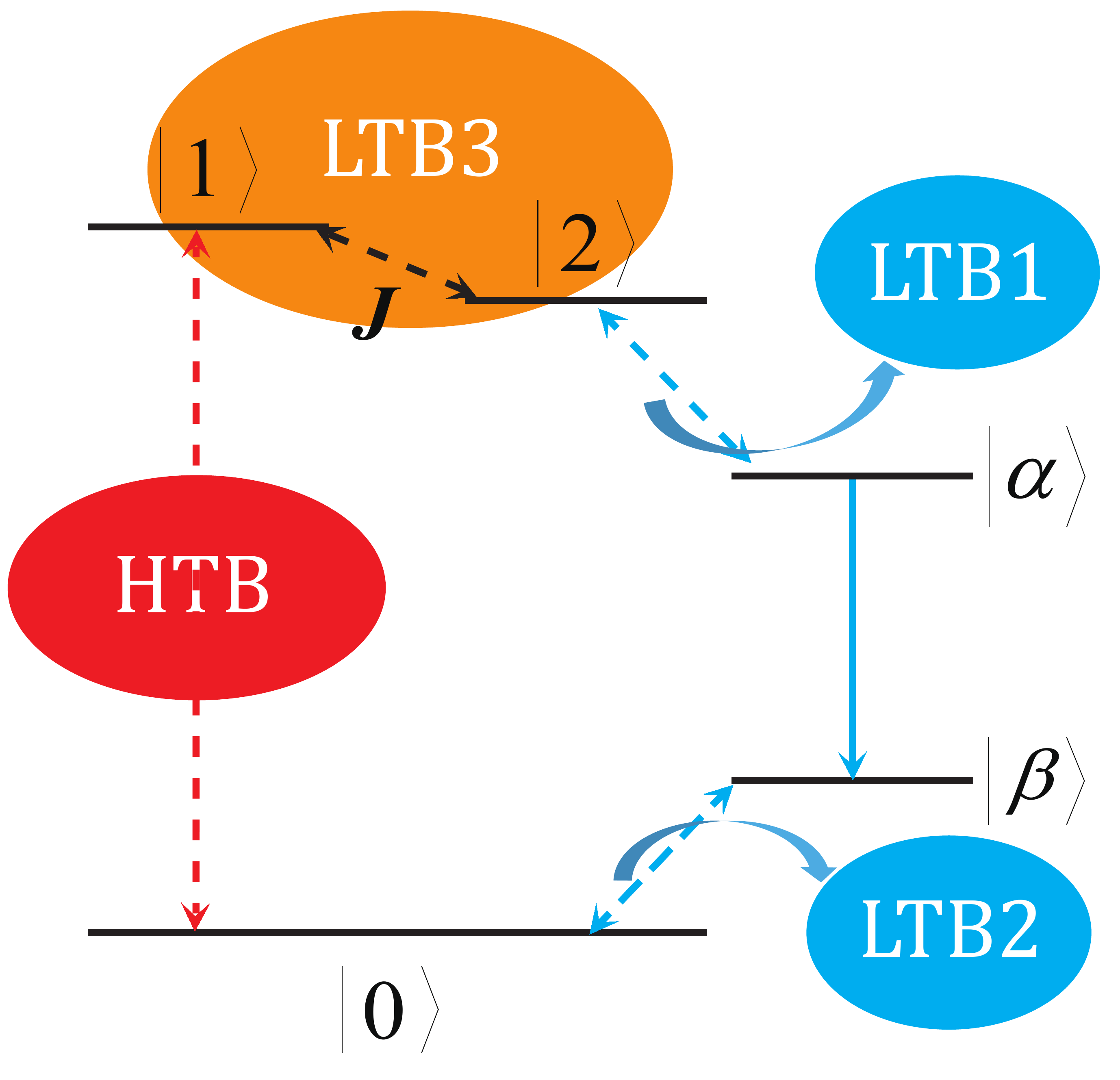}
\caption{(Color online) Schemes of the QHE model based on
PS{\rm II} RC. HTB denotes the high-temperature
photon bath from sunlight, while LTB1, LTB2 and LTB3 stand for the low-temperature
phonon baths attributed to molecular
vibrational degrees of freedom. HTB induces transition from the ground state
$\left| 0 \right\rangle $ to the single-exciton states $\left|
1 \right\rangle $. The primary charge separation denoted by the transition between $\left|
1 \right\rangle $ and $\left| 2 \right\rangle $ with interpigment coupling $J$ is subject to
LTB3. LTB1 induces transition from $\left| 2 \right\rangle $  to the secondary charge transfer state
$\left| \alpha \right\rangle $ with the excess energy radiated as a phonon, phenomenologically representing the secondary charge transfer process  in which the positive and negative charges are spatially separated.  Then the electron is released from state $\left| \alpha \right\rangle $ resulting a current from $\left| \alpha \right\rangle $ to $\left| \beta \right\rangle $. LTB2
induces transition from the positively charged state $\left| \beta
\right\rangle $
to the ground state $\left| 0 \right\rangle $ which brings the electron back to the system with emission of a phonon with excess energy.}
\label{energystructure:}
\end{figure*}

The biologically inspired quantum heat engine (QHE) transforms high-energy thermal photon radiation into low-entropy electron flux. In Fig.~\ref{energystructure:}, we analyse the photosynthetic reaction center as a five-level scheme which models the photon absorption and charge separation events described above, using the following states: the ground state $\left| 0 \right\rangle $, the exciton state $\left| {{{\left( {{\rm{Ch}}{{\rm{l}}_{D1}}{\rm{Ph}}{{\rm{e}}_{D1}}} \right)}^*}} \right\rangle  \equiv \left| 1 \right\rangle $, the ICTS $\left| {{\rm{Chl}}_{D1}^ + {\rm{Phe}}_{D1}^ - } \right\rangle  \equiv \left| 2 \right\rangle $, the secondary charge transfer state $\left| {{\rm{P}}_{D1}^ + {\rm{Phe}}_{D1}^ - } \right\rangle  \equiv \left| \alpha  \right\rangle $ and the positively charged state $\left| {{\rm{P}}_{D1}^ + {\rm{Ph}}{{\rm{e}}_{D1}}} \right\rangle  \equiv \left| \beta  \right\rangle $.
After absorption of a solar photon, the system is promoted from the ground state $\left| 0 \right\rangle $ to the exciton state $\left| 1 \right\rangle $ with a rate ${\gamma _h}$. Then the primary charge separation takes place which channels energy to the ICTS $\left| 2 \right\rangle $. This process is subjected to a phonon bath LTB3 (the third low temperature phonon bath). Then follows the secondary charge separation in which the positive and negative charges are spatially separated. For simplicity, the secondary charge separation is treated as a dissipation transition from $\left| 2 \right\rangle $ to the secondary charge transfer state $\left| \alpha \right\rangle $ with excess energy radiated as a phonon into the phonon bath LTB1 (the first low temperature phonon bath) with a rate ${\gamma _{c1}}$. This is a phenomenological treatment without going into the details of  actual parameters of PS{\rm II} RC. Then the electron is released from the system to perform work with a relaxation rate $\Gamma$, resulting in a current from $\left| \alpha  \right\rangle $ to
the positively charged state $\left| \beta  \right\rangle $ driving a chain of chemical reactions, leading eventually
to the stable storage of solar energy. The current is thus determined by the relaxation rate $\Gamma $ and the population of $\left| \alpha  \right\rangle $,
\begin{eqnarray}
\begin{aligned}
j = e\Gamma {\rho _{\alpha \alpha }}
\end{aligned}
\label{current}
\end{eqnarray}
Finally, to complete the cycle,
we assume another population transfer process to take place, emitting a
phonon with excess energy into the phonon bath LTB2 (the second low temperature phonon bath) , bringing the electron back to the neutral
ground state $\left| 0 \right\rangle $ with a rate ${\gamma _{c2}}$.

In this paper, we focus on the effects of LTB3 phonon modes on the charge transfer process of exciton-ICTS
dimer system ($\left| 1 \right\rangle $ and $\left| 2 \right\rangle $) over a broad range. Thus the coupling of the exciton state $\left| 1 \right\rangle $ and the ICTS $\left| 2 \right\rangle $ with LTB3 will be considered rigorously. To describe energy transfer between $\left| 1 \right\rangle $ and $\left| 2 \right\rangle
$, we consider a Frenkel exciton model Hamiltonian in the single exciton manifold:
\begin{eqnarray}
\begin{aligned}
 H =& {H_s} + {H_b} + {H_{sb}}, \\
 {H_s} = &\sum\limits_{m = 1,2} {{\varepsilon _m}\left| m \right\rangle \left\langle m \right|}  + J\left( {\left| 1 \right\rangle \left\langle 2 \right| + \left| 2 \right\rangle \left\langle 1 \right|} \right), \\
 {H_b} = &\sum\limits_k {{\omega _{vk}}b_k^\dag {b_k}} , \\
 {H_{sb}} =& \sum\limits_{m = 1,2} {\sigma _m^ + \sigma _m^ - \sum\limits_k {\left( {{g_{vk,m}}b_k^\dag  + g_{vk,m}^ * {b_k}} \right)} } , \\
\end{aligned}
\label{original Hamiltonian}
\end{eqnarray}
${H_s}$ describes the Frenkel-exciton Hamiltonian where ${{\varepsilon _m}}$ is the relative site energy of the exciton state $\left| 1 \right\rangle $ and the ICTS $\left| 2 \right\rangle $ with respect to the ground state $\left| 0 \right\rangle $ whose energy is set to $0$. $J$ denotes the the electronic coupling between states $\left| 1 \right\rangle $ and $\left| 2 \right\rangle $.
${H_b}$ represents the  Hamiltonian of LTB3 with ${b_k^\dag }$ (${b_k}$) the creation/annihilation operator and ${{\omega _{vk}}}$ the frequency of the $k$th phonon mode of LTB3. ${H_{sb}}$ is the interaction Hamiltonian describing the coupling of the exciton state $\left| 1 \right\rangle $ and the ICTS $\left| 2 \right\rangle $ with the phonon bath LTB3, dominated by site energy fluctuations, with ${{g_{vk,m}}}$ the coupling strength between the $k$th phonon mode and the $m$th state.

\subsection{Polaron transformation}

The formulism of polaron transformation is first used to treat charge transfer in organic molecular crystals by Holstein and then developed by Silbey etc to consider population dynamics in EET. It assumes that the electronic excitation moves collectively with its surrounding bath deformation rather than separating the exciton and bath.
Following Grover and Silbey \cite{Grover197154}, we move into the polaron frame defined by a unitary transformation $\tilde H = {e^S}H{e^{ - S}}$, where $S = \sum\limits_{m = 1,2} {\left| m \right\rangle \left\langle m \right|\sum\limits_k {{{\left( {{g_{vk,m}}b_k^\dag  - g_{vk,m}^ * {b_k}} \right)} \mathord{\left/
 {\vphantom {{\left( {{g_{vk,m}}b_k^\dag  - g_{vk,m}^ * {b_k}} \right)} {{\omega _{vk}}}}} \right.
 \kern-\nulldelimiterspace} {{\omega _{vk}}}}} } $. Within this transformed frame, the
renormalized Hamiltonian reads
\begin{eqnarray}
\begin{aligned}
 \tilde H =& {{\tilde H}_s} + {{\tilde H}_b} + {{\tilde H}_{sb}}, \\
 {{\tilde H}_s} =& \sum\limits_{m = 1,2} {{{\tilde \varepsilon }_m}\left| m \right\rangle \left\langle m \right|}  + J\kappa \left( {\left| 1 \right\rangle \left\langle 2 \right| + \left| 2 \right\rangle \left\langle 1 \right|} \right), \\
 {{\tilde H}_b} =& \sum\limits_k {{\omega _{vk}}b_k^\dag {b_k}} , \\
 {{\tilde H}_{sb}} = &J\left( {\tilde B\left| 1 \right\rangle \left\langle 2 \right| + {{\tilde B}^\dag }\left| 2 \right\rangle \left\langle 1 \right|} \right), \\
\label{renormalized Hamiltonian}
\end{aligned}
\end{eqnarray}
where ${{\tilde \varepsilon }_m}$ is the shifted on-site energy with the corresponding site-dependent reorganization energy ${\lambda _m} = \sum\limits_k {\frac{{{{\left| {{g_{vk,m}}} \right|}^2}}}{{{\omega _{vk}}}}} $, such that ${{\tilde \varepsilon }_m} = {\varepsilon _m} - {\lambda _m}$. $\tilde B$ signifies the shifted
bath operator defined as $\tilde B = B - \kappa $ with the bath operator $B = {e^{\sum\nolimits_k {{{\left( {\delta {g_{vk,12}}b_k^\dag  - \delta g_{vk,12}^ * {b_k}} \right)} \mathord{\left/
 {\vphantom {{\left( {\delta {g_{vk,12}}b_k^\dag  - \delta g_{vk,12}^ * {b_k}} \right)} {{\omega _{vk}}}}} \right.
 \kern-\nulldelimiterspace} {{\omega _{vk}}}}} }}$.  The factor $\delta {g_{vk,12}} = {g_{vk,1}} - {g_{vk,2}}$ is the difference of system-bath couplings ${{g_{vk,1}}}$ and ${{g_{vk,2}}}$ in state $\left| 1 \right\rangle $ and $\left| 2 \right\rangle $, respectively. Here, we define bath-induced renormalization factor as $\kappa  = \left\langle B \right\rangle$.
We see that, in the polaron theory, the electronic system-plus-phonon bath Hamiltonian is transformed into the polaron frame in which electronic couplings are renormalized and fluctuate due to the interaction with bath modes, while the free Hamiltonian of phonon bath LTB3 remains unchaged. In this formulism, the reorganization energy of the $m$th site can be calculated as
\begin{eqnarray}
\begin{aligned}
{\lambda _m} = \sum\limits_k {\frac{{{{\left| {{g_{vk,m}}} \right|}^2}}}{{{\omega _{vk}}}}}  = \int_0^\infty  { \frac{{{J_{mm}}\left( \omega  \right)}}{\omega }{\rm{d}}\omega},
\label{kappa}
\end{aligned}
\end{eqnarray}
We define an effective electronic couplings $\tilde J = J\kappa $, and the expectation value of the bath operator for a harmonic oscillator bath in thermal equilibrium evaluates to
\begin{eqnarray}
\begin{aligned}
 \kappa  =& \left\langle B \right\rangle  = Tr[{{\rho '}_b}B] \\
= &\exp \{  - \frac{1}{2}\int_0^\infty  {\frac{1}{{{\omega ^2}}}} [{J_{11}}\left( \omega  \right) - 2{J_{12}}\left( \omega  \right) \\
 &+ {J_{22}}\left( \omega  \right)]\coth \left( {\frac{{\beta \omega }}{2}} \right)d\omega \}
\label{kappa}
\end{aligned}
\end{eqnarray}
where $\beta  = {1 \mathord{\left/
 {\vphantom {1 {{k_B}T}}} \right.
 \kern-\nulldelimiterspace} {{k_B}T}}$ is the inverse temperature and ${{\rho '}_b} = {{\exp \left( { - \beta {{\tilde H}_b}} \right)} \mathord{\left/
 {\vphantom {{\exp \left( { - \beta {{\tilde H}_b}} \right)} {{\rm{Tr}}\left[ {\exp \left( { - \beta {{\tilde H}_b}} \right)} \right]}}} \right.
 \kern-\nulldelimiterspace} {{\rm{Tr}}\left[ {\exp \left( { - \beta {{\tilde H}_b}} \right)} \right]}}$ denotes the thermal state of the phonon bath. $\kappa $ is also called Franck-Condon factor which describes the overlap of the phonon wavefunctions, and it directly influences the effective electronic coupling ${\tilde J}$. And the spectrum functions are chosen to be super-Ohmic as
\begin{eqnarray}
\begin{aligned}
{J_{mn}}\left( \omega  \right) = {\rm{ }}{\gamma _{mn}}\frac{{{\omega ^3}}}{{{\omega _c}^2}}{e^{ - \frac{\omega }{{{\omega _c}}}}},\left( {m,n = 1,2} \right), \\
\label{spectral density function}
\end{aligned}
\end{eqnarray}
where ${\gamma _{11}}$ (${\gamma _{22}}$) signifies the dimensionless exciton (ICTS)-phonon coupling strength and ${\gamma _{12}}$ measures couplings of system and bath shared between the two states $\left| 1 \right\rangle $ and $\left| 2 \right\rangle $. As $\left| 1 \right\rangle $ and $\left| 2 \right\rangle $ refer to the exciton state and the ICTS respectively, hereafter, ${\gamma _{11}}$ and ${\gamma _{22}}$ are both named system-bath coupling strength for brevity. Note that the interaction Hamiltonian is dominated by site energy fluctuations and
$\left| 1 \right\rangle $ and $\left| 2 \right\rangle $ interact with a common bath, therefore
the energy fluctuations on the two states can have cross correlations. Consequently, the dimensionless ${\gamma _{12}}$ is utilized to to characterize the correlated fluctuations, whose amplitude is always smaller than that of the total fluctuations on each state, i.e., ${\gamma _{12}} \le \sqrt {{\gamma _{11}}{\gamma _{22}}}$. Thus we define a cross-correlation coefficient to describe bath correlation effects:
\begin{eqnarray}
\begin{aligned}
c = \frac{{{\gamma _{12}}}}{{\sqrt {{\gamma _{11}}{\gamma _{22}}} }}.
\label{cross-correlation efficient}
\end{aligned}
\end{eqnarray}
$c =  - 1,0,1$ corresponds to fully anti-correlated bath, independent(uncorrelated) bath and fully correlated bath, respectively. For numerical simulations, we assume that the two states interact with the phonon bath via the same spectral density, i.e., ${\gamma _{11}} = {\gamma _{22}} = \gamma $. In this condition, the dynamics of a single excitation is fully decoupled from the bath for the case of $\gamma =0$ or fully correlated bath.
\subsection{Polaron master equation}
After the polaron transformation $\tilde H = {e^S}H{e^{ - S}}$, the renormalized ${{\tilde H}_{sb}}$ can be taken as a perturbation term since the thermal average $\left\langle {{{\tilde H}_{sb}}} \right\rangle {\rm{ = }}0$. With the Born-Markov approximation, the polaron master equation for charge transfer process between $\left| 1 \right\rangle $ and $\left| 2 \right\rangle $ can be obtained
\begin{eqnarray}
\begin{aligned}
\frac{{{\rm{d}}{{\rho '}_s}(t)}}{{{\rm{d}}t}} =&  - i\left[ {{{\tilde H}_s},{{\rho '}_s}(t)} \right] - \sum\limits_{i,j = z, \pm } {[\Gamma _{ij}^ + {\tau _i}{\tau _j}{{\rho '}_s}(t)}\\
&  + \Gamma _{ji}^ - {{\rho '}_s}(t){\tau _j}{\tau _i} - \Gamma _{ji}^ - {\tau _i}{{\rho '}_s}(t){\tau _j} - \Gamma _{ij}^ + {\tau _j}{{\rho '}_s}(t){\tau _i}],
\label{masterequation}
\end{aligned}
\end{eqnarray}
where $\Gamma _{ij}^ \pm $ are time-dependent rates related to bath correlation function. We define ${{{\rho '}_s}(t)}$ as the reduced system density matrix in the polaron frame and define a new set of Pauli operators as
\begin{eqnarray}
\begin{aligned}
{\tau _ + } = &\left|  +  \right\rangle \left\langle  -  \right|,\\
{\tau _ - } = &\left|  -  \right\rangle \left\langle  +  \right|, \\
{\tau _z} =& \left|  +  \right\rangle \left\langle  +  \right| - \left|  -  \right\rangle \left\langle  -  \right|, \\
\label{Paulioperators}
\end{aligned}
\end{eqnarray}
where we have moved into the renormalized exciton basis in which ${{\tilde H}_s}\left|  \pm  \right\rangle  = {\varepsilon _ \pm }\left|  \pm  \right\rangle$. In appendix A, we show the detailed derivations of
Eq. (\ref{masterequation}).

In this paper, we are interested in how the phonon modes of LTB3 affects the charge transfer process of exciton-ICTS system. For the other three baths coupled with the system: HTB, LTB1 and LTB2, the Hamiltonians are respectively given by
\begin{eqnarray}
\begin{aligned}
 {H_h} =& \sum\limits_k {{\omega _{hk}}a_{hk}^\dag {a_{hk}} + } \left( {{g_{hk}}a_{hk}^\dag \left| 0 \right\rangle \left\langle 1 \right| + g_{hk}^ * \left| 1 \right\rangle \left\langle 0 \right|{a_{hk}}} \right), \\
 {H_{c1}} = &\sum\limits_k {{\omega _{c1k}}a_{c1k}^\dag {a_{c1k}} + } \left( {{g_{c1k}}a_{c1k}^\dag \left| \alpha  \right\rangle \left\langle 2 \right| + g_{c1k}^ * \left| 2 \right\rangle \left\langle \alpha  \right|{a_{c1k}}} \right), \\
 {H_{c2}} = &\sum\limits_k {{\omega _{c2k}}a_{c2k}^\dag {a_{c2k}} + } \left( {{g_{c2k}}a_{c2k}^\dag \left| 0 \right\rangle \left\langle \beta  \right| + g_{c2k}^ * \left| \beta  \right\rangle \left\langle 0 \right|{a_{c2k}}} \right), \\
\label{dissipationHamiltonian}
\end{aligned}
\end{eqnarray}
where ${a_{hk}^\dag }(i = h,c1,c2)$ is the creation operator of $i$th bath mode with the eigenfrequency ${\omega _{ik}}$ and ${{g_{ik}}}$ the corresponding coupling constant with the system.
We adopt the local Liouville operator to phenomenologically describe the corresponding dissipative transition processes \cite{Lindblad1967,Scully1997}:
\begin{eqnarray}
\begin{aligned}
{L_i}\left( \rho  \right) =& \frac{{{\gamma _i}}}{2}[({n_i} + 1)\left( {2O_i^ - \rho O_i^ +  - \left\{ {O_i^ + O_i^ - ,\rho } \right\}} \right) \\
&+ {n_i}\left( {2O_i^ + \rho O_i^ -  - \left\{ {O_i^ - O_i^ + ,\rho } \right\}} \right)],
\label{Lindblad dissipation}
\end{aligned}
\end{eqnarray}
where $i = h,c1,c2$ corresponds to the high temperature photon bath, the first and second low temperature phonon bath, respectively, ${{\gamma _i}}$ and ${n_i}$ are the corresponding dissipative rate and average photon(phonon) number. The system operators are defined as $O_h^ +  = \left| 1 \right\rangle \left\langle 0 \right|,O_{c1}^ +  = \left| 2 \right\rangle \left\langle \alpha  \right|$ and $O_{c2}^ +  = \left| \beta  \right\rangle \left\langle 0 \right|$ respectively.

Superoperator ${L_\Gamma } $ describes the process that the electron is released from the system to drive a chain of sequence chemical reactions, i.e., to perform work:
\begin{eqnarray}
\begin{aligned}
{L_\Gamma }\left( \rho  \right) = \frac{\Gamma }{2}[\left| \beta  \right\rangle \left\langle \alpha  \right|\rho \left| \alpha  \right\rangle \left\langle \beta  \right| - \rho \left| \alpha  \right\rangle \left\langle \alpha  \right| - \left| \alpha  \right\rangle \left\langle \alpha  \right|\rho ],
\label{Lindblad Gamma}
\end{aligned}
\end{eqnarray}
It leads to the electronic current proportional to the relaxation rates $\Gamma$.
\subsection{Definitions of effective voltage and power}
\begin{table}[h]
\centering \caption{Parameters used in the numerical simulations.}
\includegraphics[scale=0.426]{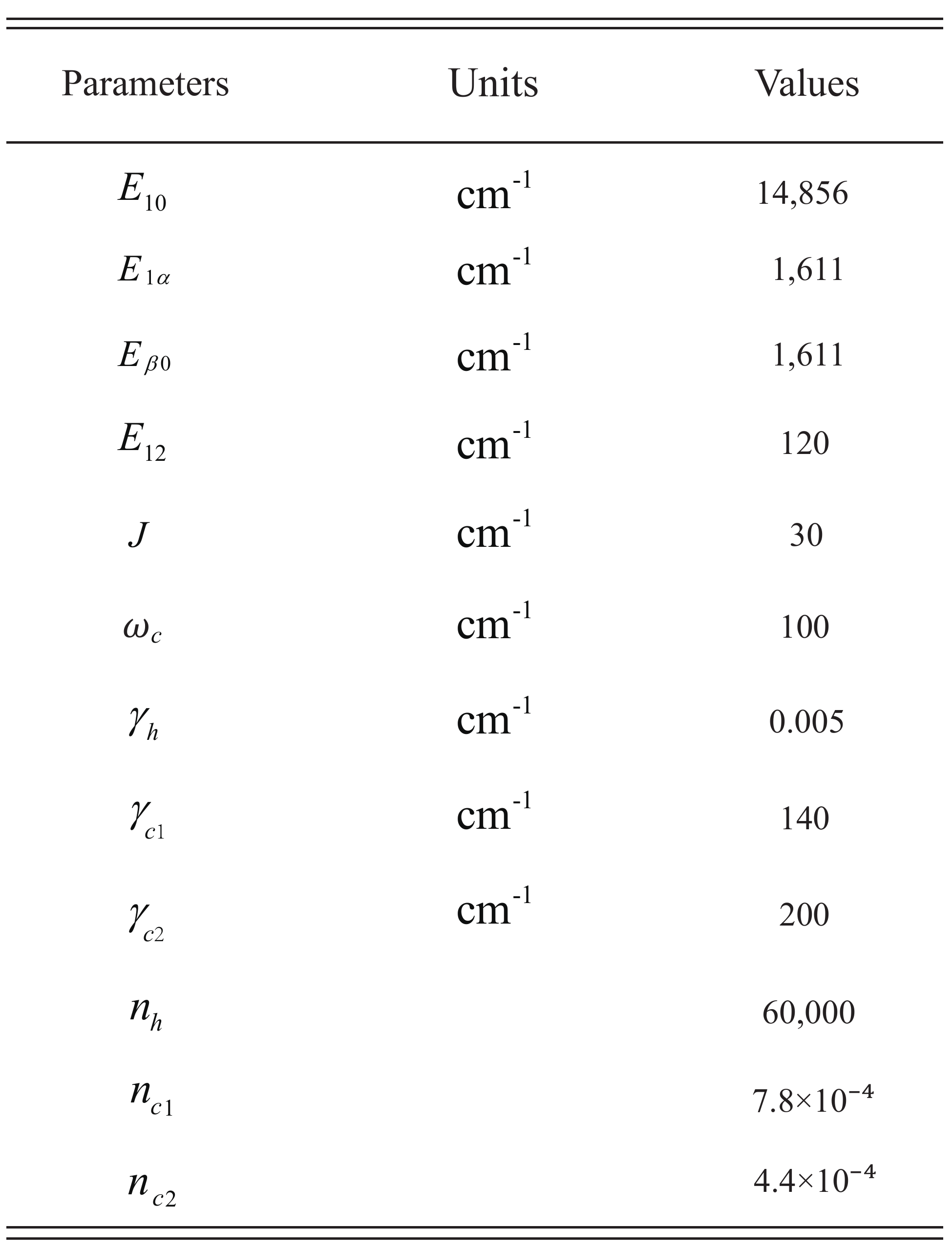}
\label{table:}
\end{table}
In the reaction center, secondary charge separation and the following electron's being released to perform work induce a current which can be thought to flow cross a load connecting $\left| \alpha
\right\rangle $ and $\left| \beta  \right\rangle $. Introducing an effective voltage $V$ as a drop of the electrostatic potential across the load, we yield
\begin{eqnarray}
\begin{aligned}
eV = {E_\alpha } - {E_\beta } + {k_B}T\ln \frac{{{\rho _{\alpha \alpha }}}}{{{\rho _{\beta \beta }}}}
\label{voltage}
\end{aligned}
\end{eqnarray}
for our model, where $e$ is the electric charge, ${E_i}$ is the energy of state $\left| i \right\rangle $ and ${\rho _{ii}}$ is the population of state $\left| i \right\rangle (i = \alpha ,\beta )$. With the current (Eq. (\ref{current})) and voltage, the power output can be calculated as
\begin{eqnarray}
\begin{aligned}
P = j \cdot V.
\label{power}
\end{aligned}
\end{eqnarray}
The performance of our QHE can be assessed in terms of the photovoltaic properties of PS{\rm II} RC complex, i.e., the
steady-state current-voltage ($j - V$) and power-voltage ($P - V$)
characteristics. Using the steady-state solution, we plot the $j - V$  curve and power at increasing rate
$\Gamma $, while keep the other parameters fixed: $\Gamma
\to 0$ ($j \to 0$) corresponds  to the open-circuit case, and in the
short-circuit case,  $V \to 0$.

The parameters used in numerical simulations are listed in Table ${\rm I}$.
These parameters are reported in recent literatures \cite{Dorfman2013110,Abramavicius2010133185401,Madjet2006110,Abramavicius2010133}
and they are used in the simulation in
\cite{Dorfman2013110,Creatore2013111,Xu201490,Qin2016144}. The energy
differences are defined as ${E_{ij}} = {E_i} - {E_j}$.

\section{Results and discussions}

In the polaron theory, the unitary transformation effectively changes the basis to the polaron basis
including the exciton/charge transfer states and their associated displaced phonon modes. The thermal-averaged
renormalized electronic
coupling incorporates temperature and system-bath coupling dependence into the zero-order transformed
Hamiltonian to evaluate
the effects of thermal fluctuations on the eigenstates and equilibrium structures of the exciton-ICTS dimer
system. We can predict that the renormalized electronic coupling goes to $0$ at large temperature and
system-bath couplings. In this section, we investigate how the temperature and system-bath coupling
strength affect the effective coupling ${\tilde J}$ (or Franck-Condon factor $\kappa$), coherence and
dynamical localization of the exciton-ICTS dimer system, and therefore, $j-V$ and $P-V$ characteristics as well.

\subsection{Effective electronic coupling, coherence and delocalization length}

\begin{figure}[htbp]
\centering
\includegraphics[scale=0.3]{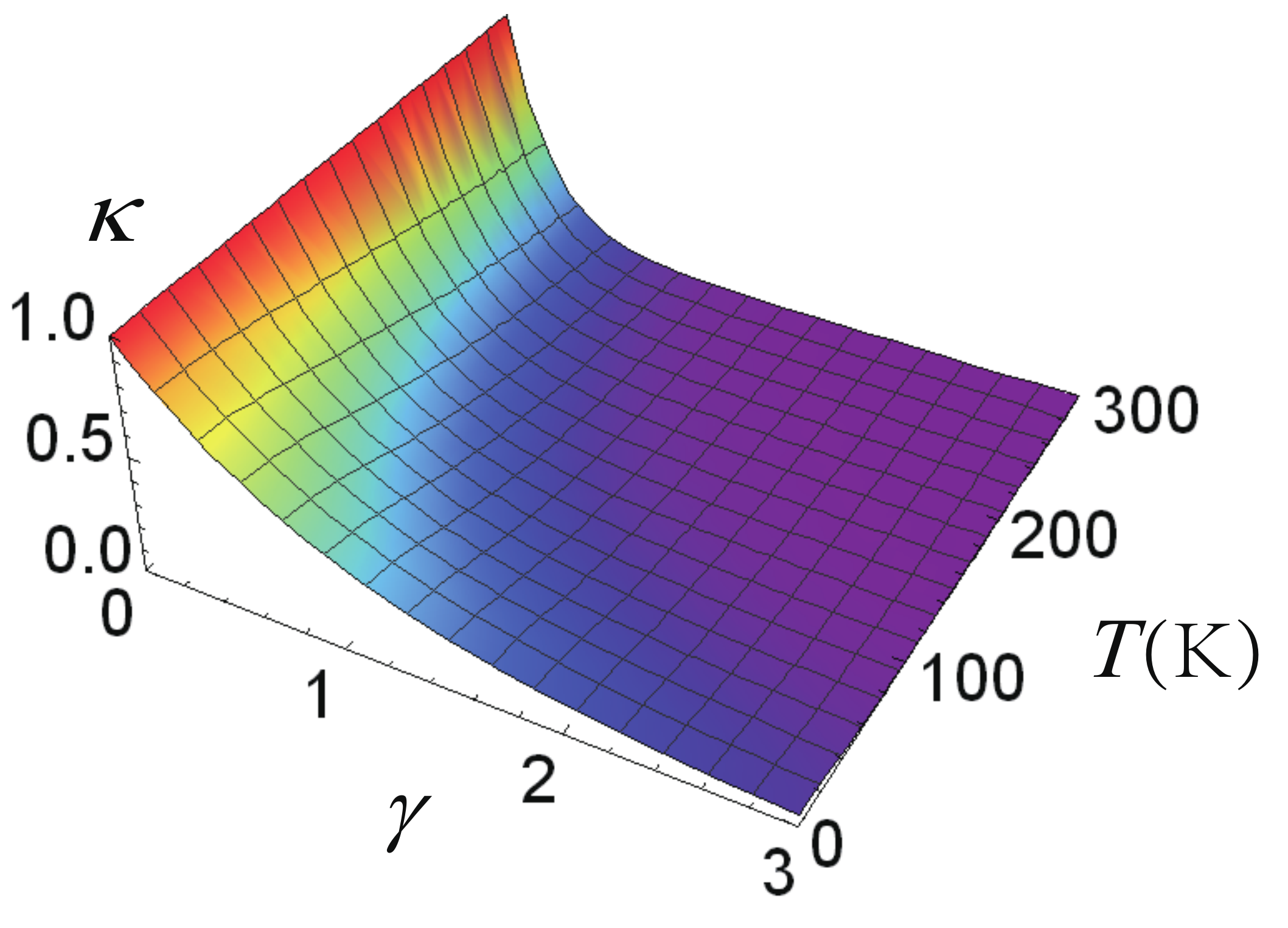}
\caption{(Color online) Franck-Condon factor $\kappa$ as a function of temperature $T$ and system-bath coupling $\gamma $ with cross-correlation coefficient $c=0$.}
\label{kappa:}
\end{figure}

Fig.~\ref{kappa:} exhibits the effects of temperature $T$ and system-bath coupling strength $\gamma $ on
Franck-Condon factor $\kappa$. We see that $\kappa$ diminish to $0$ when either $T$ or $\gamma $ increases.
From Eq. (\ref{kappa}), we can also conclude that $\kappa $ decays to $0$ at high temperature or strong
system-bath coupling. This leads to decreasing effective electronic coupling $\tilde J$ at either of these
two limits and therefore, influences the performance of QHE which will be discussed in details in the
following section.
\begin{figure}[htbp]
\centering
\includegraphics[scale=0.32]{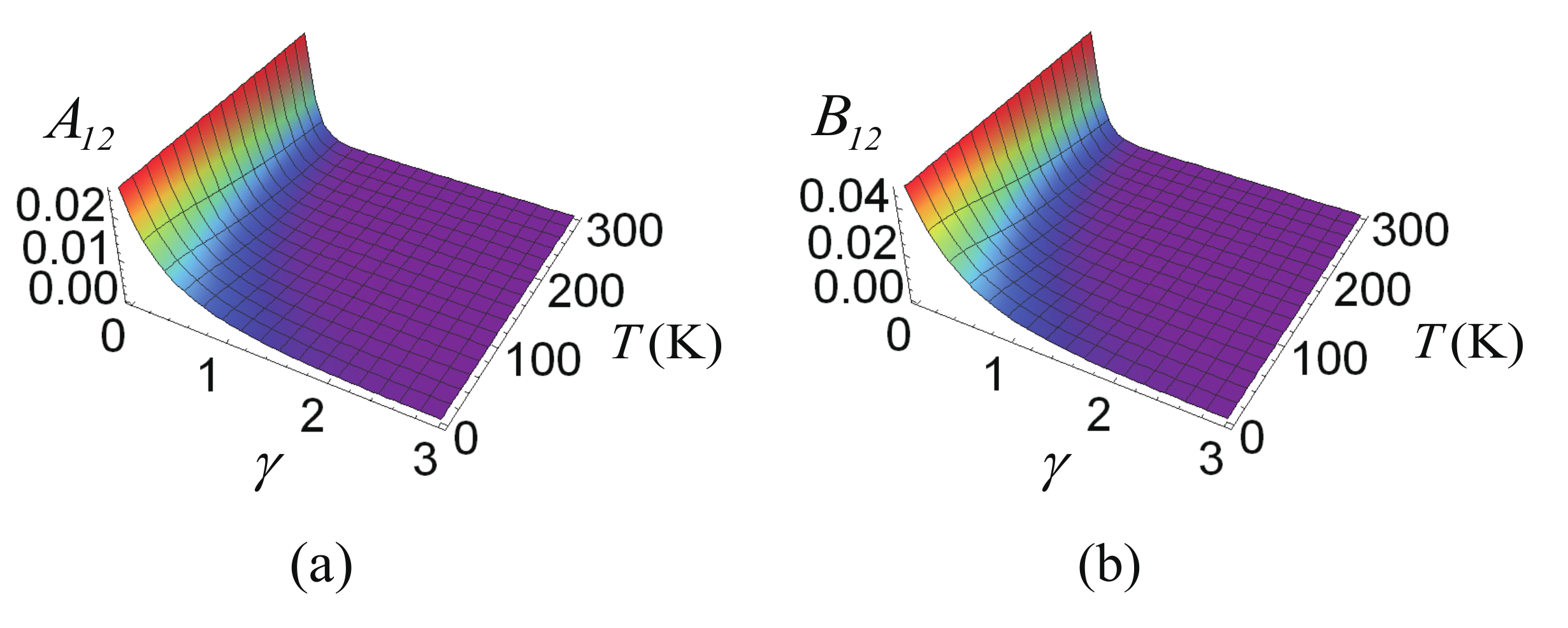}
\caption{(Color online) Coherence of the exciton-ICTS dimer system as a function of temperature $T$ and system-bath coupling $\gamma $. (a) Real part of off-diagonal density matrix element: ${A_{12}} = {\mathop{\rm Re}\nolimits} \left[ {{\rho _{12}}} \right]$. (b) Imaginary part of off-diagonal density matrix element: ${B_{12}} = {\mathop{\rm Im}\nolimits} \left[ {{\rho _{12}}} \right]$. The cross-correlation coefficient $c=0$.}
\label{coherence:}
\end{figure}
The equilibrium structure of the exciton-ICTS dimer system are also affected by the temperature and
system-bath coupling strength dependence of the effective coupling $\tilde J$ (or $\kappa$) as shown in Fig.~\ref{coherence:}. The coherence
elements ${A_{12}}$ and ${B_{12}}$ of the equilibrium density matrix are illustrated in
Fig.~\ref{coherence:} (a) and (b) respectively, from which we conclude that the coherence decays to zero
with increasing temperature or system-bath coupling strength. Therefore, strong system-bath
coupling deteriorates the coherence of the exciton-ICTS dimer system. This is owing to dynamical energy
fluctuations induced by the phonon bath. To explore the role of temperature and system-bath coupling
strength in the system equilibrium structure, we also adopt the concept of delocalization length $L$, defined
as the inverse participation of the eigenstate ($\left|  +  \right\rangle $ or $\left|  -  \right\rangle $ in our model):
\begin{eqnarray}
\begin{aligned}
L = \frac{1}{{{{\left| {\left\langle 1 \right|\left.  \pm  \right\rangle } \right|}^4} + {{\left| {\left\langle 2 \right|\left.  \pm  \right\rangle } \right|}^4}}}.
\label{delocalizationlength1}
\end{aligned}
\end{eqnarray}
In the renormalized exciton basis,
${\tilde H_s}\left|  \pm  \right\rangle  = {\varepsilon _ \pm }\left|  \pm  \right\rangle $ with
\begin{eqnarray}
\begin{aligned}
 \left|  +  \right\rangle  = \cos \frac{\theta }{2}\left| 1 \right\rangle  + \sin \frac{\theta }{2}\left| 2 \right\rangle , \\
 \left|  -  \right\rangle  = \sin \frac{\theta }{2}\left| 1 \right\rangle  - \cos \frac{\theta }{2}\left| 2 \right\rangle , \\
\label{eigenstate}
\end{aligned}
\end{eqnarray}
(see appendix A Eq. (\ref{newbasis})). Consequently, delocalization length
$L$ is given by
\begin{eqnarray}
\begin{aligned}
L = \frac{1}{{{{\left| {\cos \frac{\theta }{2}} \right|}^4} + {{\left| {\sin \frac{\theta }{2}} \right|}^4}}}
  \label{delocalizationlength2}
\end{aligned}
\end{eqnarray}

\begin{figure}[htbp]
\centering
\includegraphics[scale=0.4]{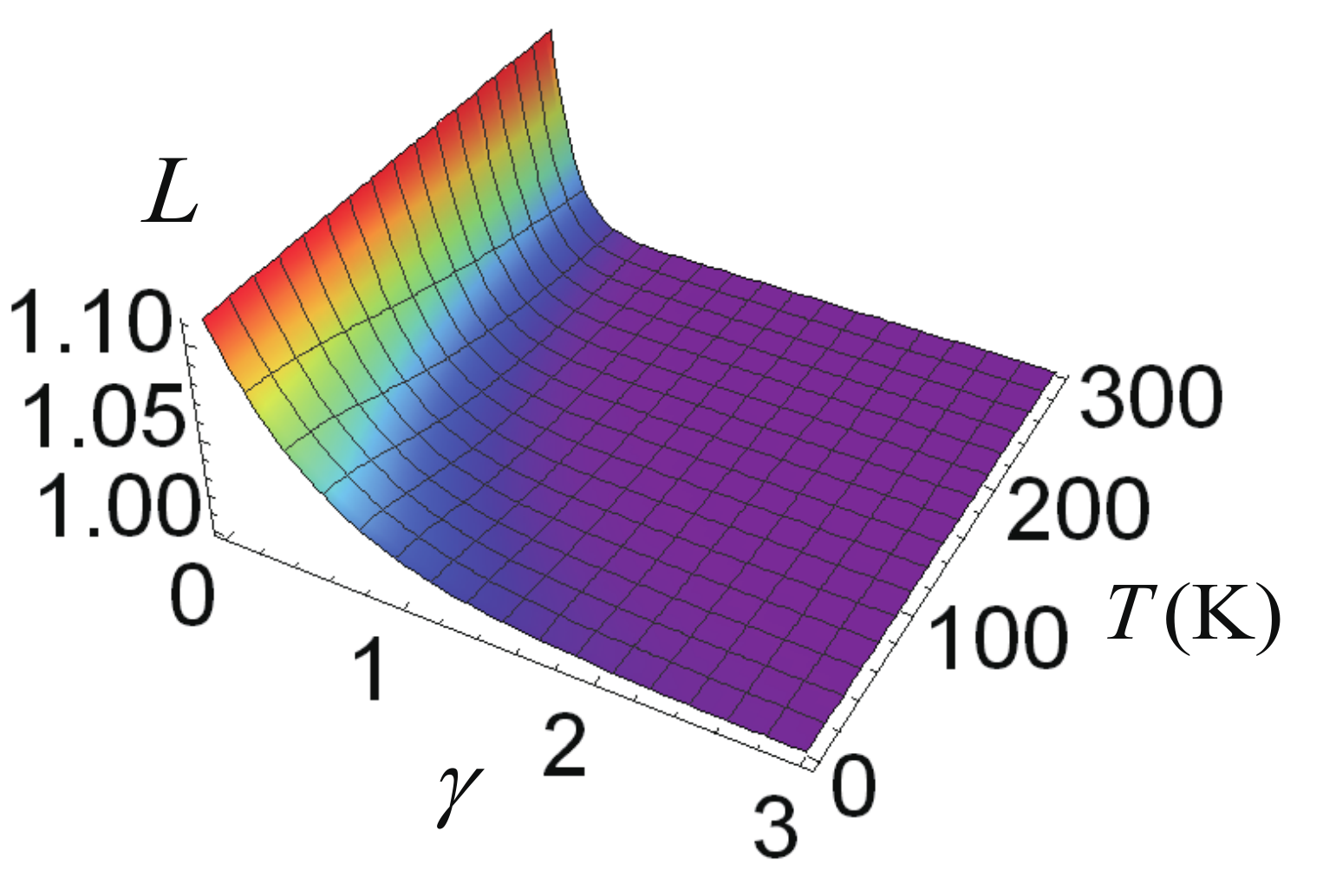}
\caption{(Color online) Delocalization length of the exciton-ICTS dimer system as a function of temperature $T$ and system-bath coupling $\gamma $. The cross-correlation coefficient $c=0$.}
\label{delocalizationlength:}
\end{figure}

In Fig.~\ref{delocalizationlength:}, we plot the delocalization length as a function of temperature $T$ and
system-bath coupling strength $\gamma $. It is observed that, as $T$ or $\gamma $ increases, the
delocalization length decreases to $1$, which means eigenstates are completely localized on $\left| 1 \right\rangle $ and $\left| 2 \right\rangle $. This also can be attributed to the bath-induced energy
fluctuations. We see that in the polaron theory, this localization effect manifested by the renormalization
of the effective electronic coupling $\tilde J$ is embodied in the zero-order transformed Hamiltonian which
is an advantage of this approach. In the following discussion, we apply dynamical localization to
characterize the impact of phonon bath, giving explanations of various dynamical behaviors.

\subsection{Steady-state j-V and P-V characteristics}

\begin{figure}[htbp]
\centering
\includegraphics[scale=0.3]{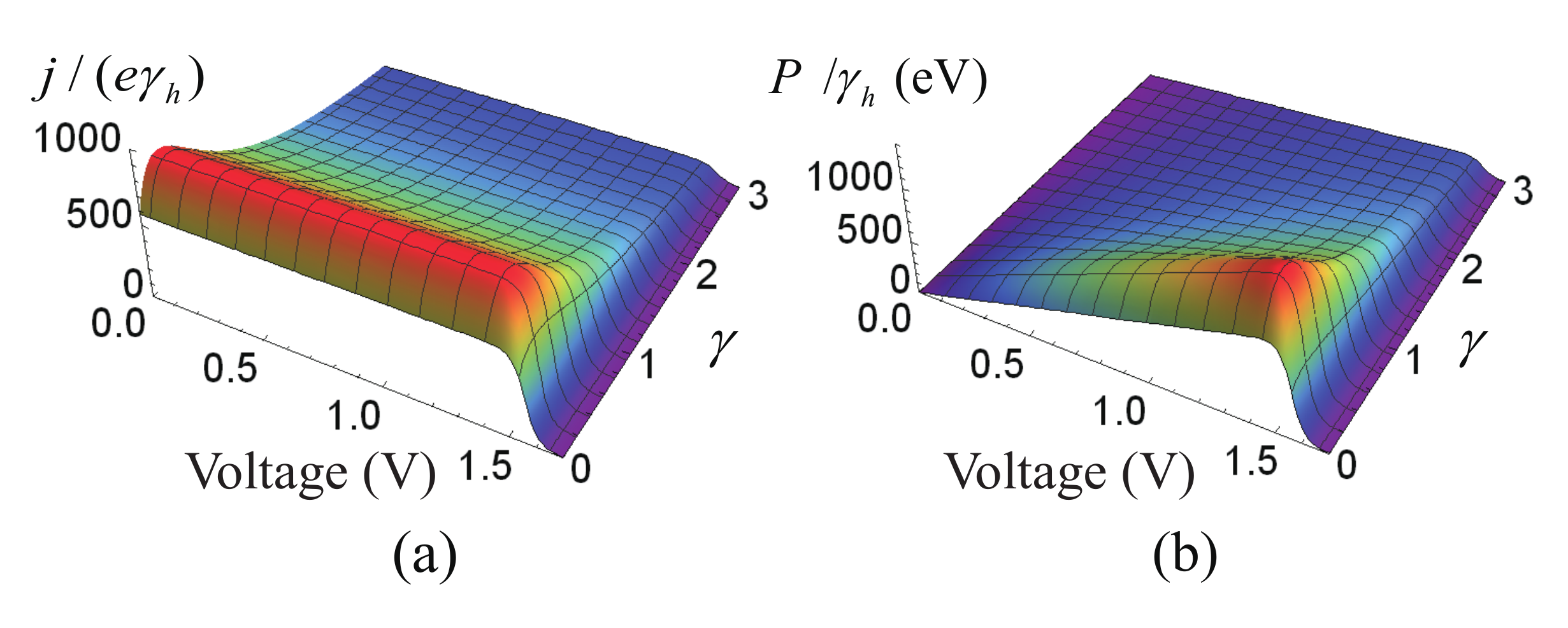}
\caption{(Color online) Steady-state $j-V$ (a) and $P-V$ (b) characteristics of QHE as a function of system-bath coupling strength $\gamma $ with cross-correlation coefficient $c = 0$ and temperature $T = 300{\rm{K}}$.}
\label{couplingstrength:}
\end{figure}

Fig.~\ref{couplingstrength:} (a) and (b) show steady-state $j-V$ and $P-V$ characteristics respectively, as a function of system-bath coupling strength $\gamma $.
At weak system-bath couplings, the current and power increases with $\gamma $. This can be explained from
the point of view of noise-assisted transport extensively reported in the previous
works \cite{Rebentrost200911,Caruso2009131,Chin201012,Wu201012,Sarovar201183}. The range of on-site energies of the two states are broadened due to bath-induced energy fluctuations, which leads to the overlap of $\left| 1 \right\rangle $ and $\left| 2 \right\rangle $
in energy and thus increases transfer rate between the two states enhancing steady-state $j-V$ and $P-V$ characteristics of QHE. If $\gamma $ continues to increase, the
effect of resonant mode decreases as the energy of each state is
distributed over a very large interval. And meanwhile, increasing system-bath coupling strength $\gamma $
also deteriorates the coherence of the exciton-ICTS dimer system due to bath-induced fluctuations, or
equivalently, leads to dynamical localization that the electronic eigenstates are localized on the state
$\left| 1 \right\rangle $ or $\left| 2 \right\rangle $ which impedes charge transfer process and thus
lessens the current and power generated.
Thus, there exists an optimal value of $\kappa$ at which noise-assisted effect and dynamical localization
reach a balance producing the maximal current and power. Interestingly, these two effects both originate
from the fluctuations induced by phonon bath. The fluctuations can promote or impede the current and power
according to the values of $\gamma$.

\begin{figure}[htbp]
\centering
\includegraphics[scale=0.17]{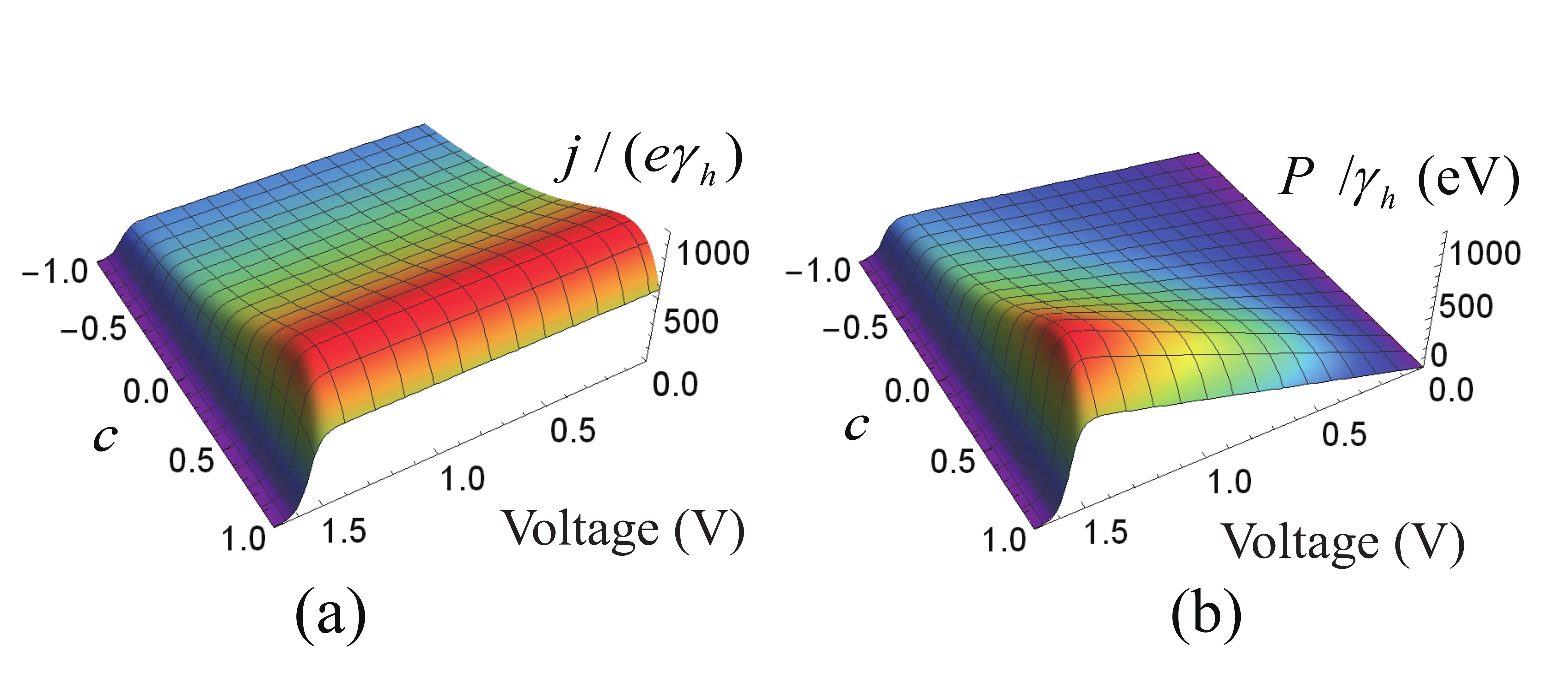}
\caption{(Color online) Steady-state $j-V$ (a) and $P-V$ (b) characteristics of QHE as a function of bath cross-correlation coefficient $c$ (defined as Eq. (\ref{cross-correlation efficient})) with system-bath coupling strength $\gamma=1$ and temperature $T = 300{\rm{K}}$.}
\label{crosscorrelation:}
\end{figure}

In Fig.~\ref{couplingstrength:}, we set the cross-correlation coefficient $c = 0$, i.e., the bath is uncorrelated. Recent experiments have revealed that, besides electronic coherence, highly correlated fluctuations of the bath should also be included to explain the observed long-lasting excitonic coherence. Motivated by these findings, the effect of bath correlations on the dynamics of EET
in photosynthetic light-harvesting complexes has attracted great interest recently \cite{McCutcheon201183,Chang2012137,Sarovar201183,Chen2010132,Fassioli20101,Struempfer2011134,Olbrich2011115,Jing2012116}. Next we shed light on the effects of bath correlation.

In Fig.~\ref{crosscorrelation:}, we plot steady-state $j-V$ (a) and $P-V$ (b) characteristics as a function of cross-correlation coefficient $c$. Similar to the case of $\gamma $, the current or power is not a monotonic function of $c$. We comprehend this by carefully examining Eq. (\ref{kappa}), which can be analytically calculated as
\begin{eqnarray}
\begin{aligned}
\kappa  = \exp \left[ { - \frac{{\left( {{\gamma _{11}} - 2c\sqrt {{\gamma _{11}}{\gamma _{22}}}  + {\gamma _{22}}} \right)}}{2}\left( { - 1 + \frac{{\psi '\left( {\frac{1}{{\beta {\omega _c}}}} \right)}}{{{{\left( {\beta {\omega _c}} \right)}^2}}}} \right)} \right].
\label{analyticalkappa}
\end{aligned}
\end{eqnarray}
where $\psi '$ denotes the trigamma function \cite{Jeffreys1998}. We set $y = {\gamma _{11}} - 2c\sqrt {{\gamma _{11}}{\gamma _{22}}}  + {\gamma _{22}}$. Since the assumption ${\gamma _{11}} = {\gamma _{22}} = \gamma $ has been made, we have $y = 2\gamma (1 - c)$. From Fig.~\ref{crosscorrelation:}, the maximal
current and power appears when $c  \sim 0.8$ and this gives $y \sim 0.4$ with the fixed system-bath
coupling $\gamma = 1$ when temperature $T=300$ K. And from Fig.~\ref{couplingstrength:}, taking $\gamma  \sim 0.2$ yields the maximal current and power and thus $y \sim 0.4$ with the fixed cross-correlation
coefficient $c=0$ when temperature $T=300$ K, in accordance with the results of Fig.~\ref{crosscorrelation:}. Thus $y$ can be regarded as a whole which we name as effective system-bath
coupling factor. It characterizes how the exciton-ICTS dimer system interacts with bath modes, including the
effects of both individual exciton- or ICTS-phonon coupling (denoted by ${\gamma _{11}}$ and ${\gamma _{22}}$ respectively) and correlated fluctuations of bath modes (denoted by ${\gamma _{12}}$ or $c$) on system-bath interaction. For a
certain temperature, there exists an optimal value for $y$. For Fig.~\ref{couplingstrength:} and Fig.~\ref{crosscorrelation:} with fixed temperature $T = 300{\rm{K}}$, we have evaluated the optimal value
of the effective system-bath coupling strength $y$ to be $0.4$.  $y \sim 0 - 0.4$ is noise-assisted regime in which the current and power increases with $y$. At $y = 0.4$, noise-assisted effects and dynamical
localization reach a good balance generating maximal current and power. When $y > 0.4$, dynamical delocalization dominates the charge transfer process, increasing $y$ giving rise to enhanced fluctuations which hinders the QHE performance. Therefore, we conclude from $y = 2\gamma (1 -
c)$ that the effect of bath correlation $c$ on current and power is incorporated in the effective
system-bath coupling strength $y$. $c$ effectively strengthens or weakens $\gamma$ giving different values
of $y$ and thus impacts $j-V$ and $P-V$ characteristics. Now, we analyse
Fig.~\ref{couplingstrength:} in terms of effective system-bath coupling factor $y$: when $T=300$ K, for fixed $c=0$, $\gamma \sim 0 - 0.2$ is the noise-assisted transport regime ($y \sim 0 - 0.4$). If $\gamma$ increases further, $y$ locates in the dynamical localization regime where larger $\gamma$ reduces the current and power. Similarly,
Fig.~\ref{crosscorrelation:} can be reinterpreted as follows: when $c=1$, we obtain $y=0$ which corresponds to $\gamma=0$ in Fig.~\ref{couplingstrength:}. As $c$ gradually decreases, $y$ first increases to its optimal
value corresponding to the noise-assisted transport regime, and then increases into the dynamical
localization regime. Thus as $c$ decreases, the current and power increases first and then decreases. Based
on these discussions, we further inspect how $\gamma$ and cross-correlation coefficient $c$ conspire to
affect the CT process in Sec. {\rm III} C.

\begin{figure}[htbp]
\centering
\includegraphics[scale=0.15]{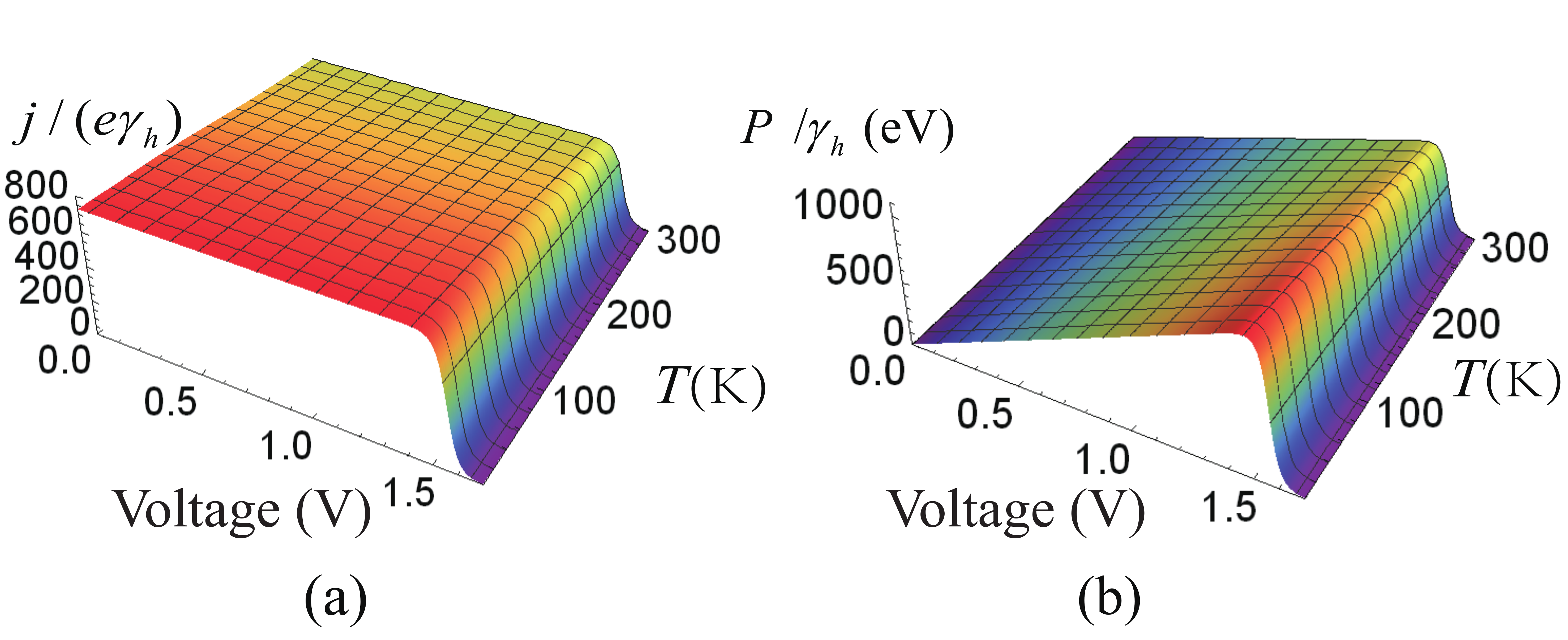}
\caption{(Color online) Steady-state $j-V$ (a) and $P-V$ (b) characteristics of QHE as a function of temperature $T$ with the
system-bath coupling strength $\gamma = 1$ and cross-correlation coefficient $c=0$.}
\label{temperature:}
\end{figure}

In Fig.~\ref{temperature:}, we plot steady-state $j-V$ and $P-V$ characteristics for varying temperature
$T$. We have learned from above discussion that for the fixed
temperature $T=300$ K, the optimal $y$ takes $0.4$. The dynamics of
the exciton-ICT dimer system includes noise-assisted and dynamical localization regime.
From Fig.~\ref{temperature:}, however, we see that as temperature increases, the current and power decrease monotonically. We can explain as follows: higher
temperature reduces effective electronic coupling $\tilde J$ from Fig.~\ref{kappa:} or Eq. (\ref{analyticalkappa}), which in turn deteriorates coherence or leads
to dynamical localization due to enhanced bath-induced fluctuations. But this raises a question that: since both higher temperature and larger $\gamma$ (or $y$) lead
to dynamical localization, does the dynamics of
exciton-ICT dimer system converts between noise-assisted and dynamical localization regime with varying $T$ just as that with $\gamma$ (or $y$) ? Or how can the effect of temperature be reflected ? Based on this question, we investigate how $\gamma$, $c$ and $T$ conspire to affect the QHE performance in Sec. {\rm III} D.

\subsection{Combined effects of $\gamma$, $c$}

\begin{figure}[htbp]
\centering
\includegraphics[scale=0.28]{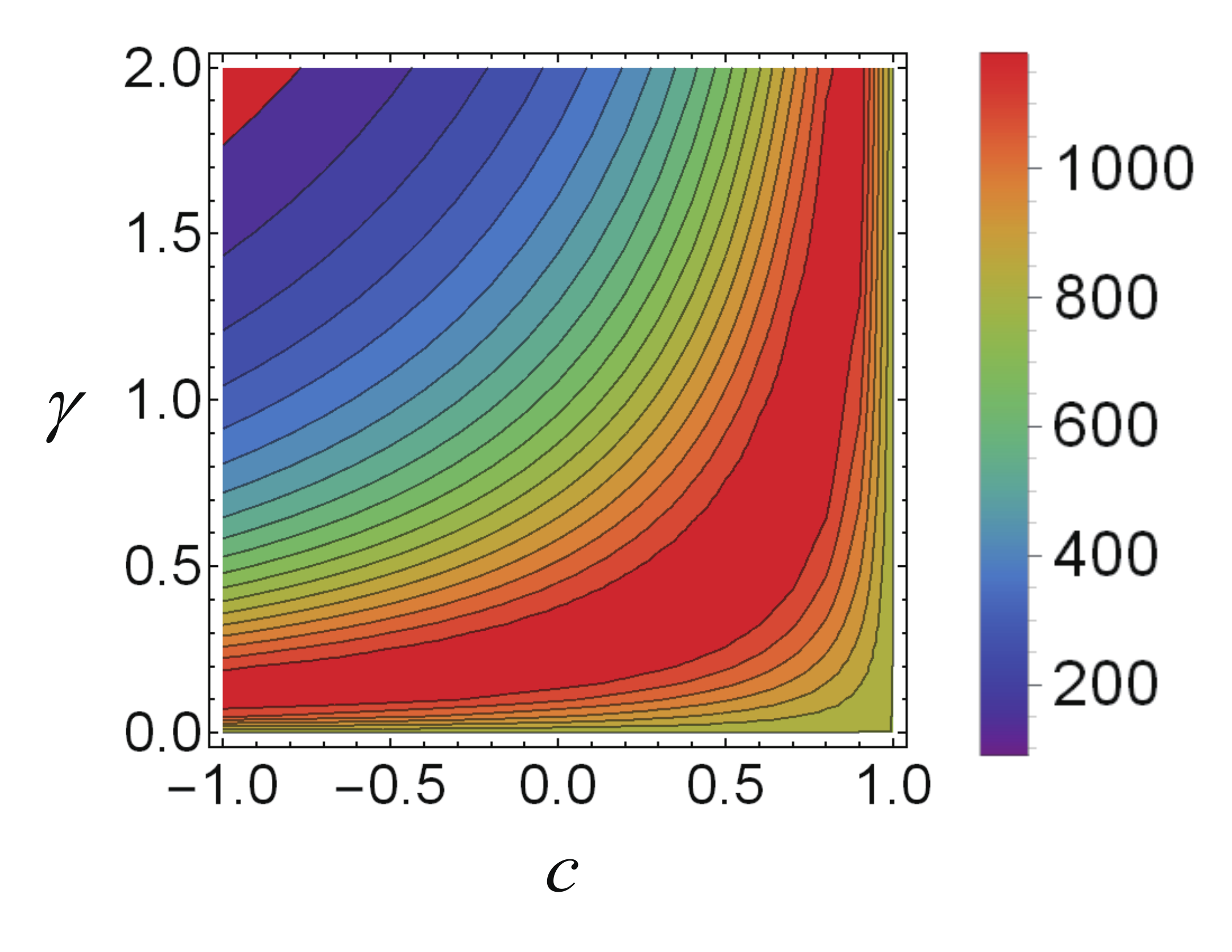}
\caption{(Color online) The  power
generated by QHE as a function of cross-correlation coefficient $c$ and system-bath coupling strength $\gamma $.}
\label{couplingstrengthcrosscorrelation:}
\end{figure}

In Fig.~\ref{couplingstrengthcrosscorrelation:}, we plot the  power of our QHE model as a function of system-bath
coupling strength and bath correlations, from weak ($\gamma  \sim 0$) to strong ($\gamma = 2$) coupling,
and from fully correlated bath ($c=1$) to fully anti-correlated bath ($c=-1$). Although, we can also plot the current, basically the variation of current is the same as that of power. Thus we only focus on the behaviors of the power generated.

The QHE performance shows
strong dependence on both $\gamma$ and $c$ and various dynamical regimes which can be interpreted in terms
of what we have learned from Eq. (\ref{analyticalkappa}) in Section {\rm III} B. Nevertheless, for weak and strong
coupling strength, the dependence of the current and power on bath correlation is remarkably different.

From Fig.~\ref{couplingstrengthcrosscorrelation:}, when $\gamma$ is very small ($ < 0.1$), i.e., at weak system-bath couplings, the power
generated by the QHE model monotonically decreases with increasing $c$. That's to say, the power reaches maximum
when the bath is fully anti-correlated. We see that when $\gamma$ is very small, the effective system-bath coupling $y$ locates in noise-assisted
transfer regime ($y \sim 0 - 0.4$), increasing $y$ leads to higher performance of QHE. And according to Eq. (\ref{analyticalkappa}) or the effective
system-bath coupling expression $y = 2\gamma (1 - c)$, $y$ is a  monotonically decreasing function of $c$, thus increasing $c$ gives smaller effective
system-bath coupling factor $y$, leading to reduced power. Also, we can learn from Sec. {\rm III} B
that, anti-correlated bath gives rise to enhanced fluctuations in the effective coupling ${\tilde J}$ while
correlated bath suppress fluctuations. As the optimal effective
system-bath coupling factor takes $y \sim 0.4$ when $T=300$ K, it is straightforward to evaluate the
optimal value of $c$ as smaller than $-1$, consequently the anti-correlated bath with enhanced fluctuations in the effective electronic coupling $\tilde J$ maximizes the QHE performance. And when $c$ increases from $-1$ to $1$, the power monotonically decreases.

However, when $\gamma$ gets larger, for example, $\gamma=1$, the
power first increases to a maximum and then decreases with increasing $c$ from
Fig.~\ref{couplingstrengthcrosscorrelation:}. According to the expression
$y = 2\gamma (1 - c)$, with $\gamma=1$, $c \sim 0.8$ obtains the optimal effective system-bath coupling
factor $y$, in agreement with what can be observed from Fig.~\ref{couplingstrengthcrosscorrelation:}.
$c \sim -1-0.8$  corresponds to dynamical
localization regime in which anti-correlated bath with enhanced fluctuations will certainly reduce the
power, while $c \sim 0.8-1$ noise-assisted regime, i.e. varying $c$ between $-1$ and $1$ obtain
a range of $y$ that contains the optimal
value $y = 0.4$. As a consequence, when $\gamma=1$, the power is not a monotonic function of
$c$, just as observed in Fig.~\ref{couplingstrengthcrosscorrelation:}. The bath-induced fluctuations promote or impede the QHE model for different values of $c$.

Next, we discuss the case when $c$ is fixed observed in Fig.~\ref{couplingstrengthcrosscorrelation:}.
Similarly, when the bath is basically correlated ($c \sim 1$), the  power monotonically
increases with $\gamma$ since when $c \sim 1$, $y$ is in the range of noise-assisted transport regime
unless $\gamma$ becomes extremely large. Increasing $y$ produces larger
power. And
$y = 2\gamma (1 - c)$ is a monotonically increasing function of $\gamma$. Therefore, increasing $\gamma$
leads to enhanced fluctuations  promoting the QHE performance. Otherwise ($c < 0.85$), when $c = 0.5$ for
example, the power first increases to the maximum and then decreases with increasing $\gamma$. As $c = 0.5$
gives $y = \gamma$, with increasing $\gamma$, $y$ increases from noise-assisted transport regime to
dynamical localization regime, passing the optimal value $y=0.4$. Therefore, when $c < 0.85$, the power is not a monotonic function of $\gamma$.

What we have discussed above is based on a fixed temperature  $T = 300{\rm{K}}$. It is clear from
Eq. (\ref{analyticalkappa}) that temperature also affects the thermal averaged Franck-Condon factor
$\kappa$. So it is interesting to study how temperature works together with system-bath coupling $\gamma$
and cross-correlation coefficient $c$ to affect QHE behaviors.

\subsection{Effects of temperature $T$}

\begin{figure}[htbp]
\centering
\includegraphics[scale=0.28]{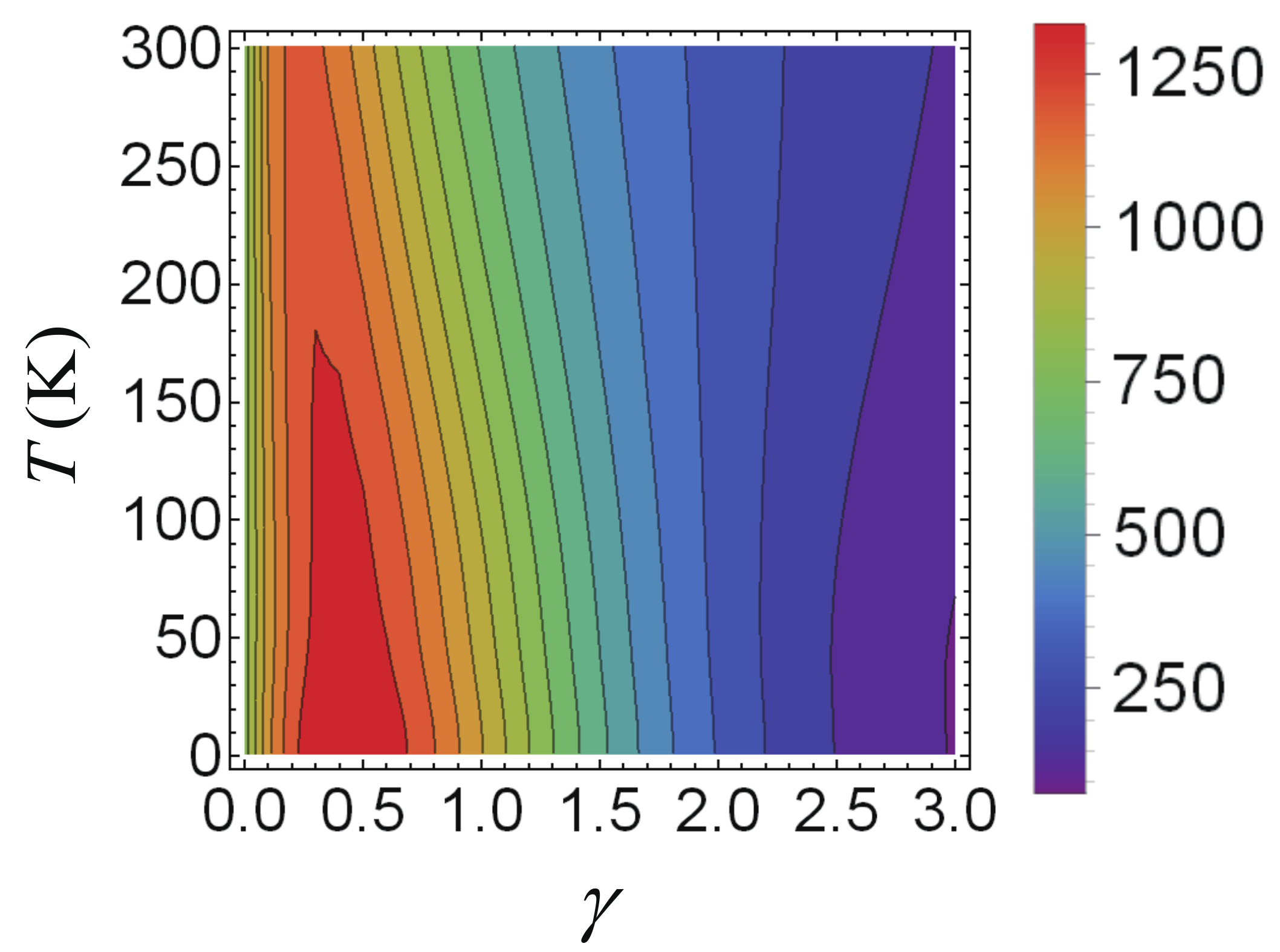}
\caption{(Color online) The power
generated by QHE as a function of temperature $T$ and system-bath coupling strength $\gamma $ for the uncorrelated bath ($c=0$).}
\label{uncorrelated:}
\end{figure}


Fig.~\ref{uncorrelated:} is plotted for the power
generated by QHE as a function of temperature $T$ and system-bath coupling strength $\gamma $ for the uncorrelated
bath ($c=0$). At any temperature, the power increases first to a maximal value
and then gradually decreases with $\gamma$, just as revealed by Fig.~\ref{couplingstrength:} and
Fig.~\ref{couplingstrengthcrosscorrelation:}. In contrast, the power decreases with temperature monotonically when $\gamma $ is not very large, i.e., the dynamics of exciton-ICTS dimer system locates in the dynamical localization regime in this case. However we can observe that, temperature affects the optimal value of
$\gamma$ that yields the maximal power. We revisit Eq. (\ref{analyticalkappa}) to give an
explanation. It is easy to know that an optimal effective system-bath coupling strength $y$ correspondingly
gives an optimal Franck-Condon factor $\kappa$ for a certain temperature $T$. The effects of phonon bath
LTR3 on the exciton-ICT dimer system are fully embodied in Franck-Condon
factor $\kappa$. Since $\tilde J = J\kappa $, $\kappa $ directly influences the effective electronic
coupling $\tilde J$ which determines the extent of delocalization for the dimer system induced by fluctuations. We can estimate the
optimal $\kappa$ (denoted by ${\kappa _{opt}}$) for different temperatures from Fig.~\ref{uncorrelated:} in
which cross-correlation coefficient $c$ are set to be zero.
For $T=300$ K, the maximal power appears at $\gamma  \sim 0.25$, this yields ${\kappa _{opt}} \sim 0.34$. Similarly, for $T=200$ K, ${\kappa _{opt}}=0.41$, and for $T=50$ K, ${\kappa _{opt}}=0.57$.
In physics, this indicates that temperature influences the balance between noise-assisted transport effects and
dynamical localization, though the dynamics can not convert between the two regimes with temperature just as that with system-bath coupling and bath-correlation.

As system-bath coupling further increases ($\gamma  > 2.2$), it is interesting to observe that the power
gets slightly larger with increasing temperature. Actually, we ignore another effect induced by temperature in the
discussions above. That is: higher temperature gives rise to larger effective transfer rates concerned with LTR1 and LTR2, which enhances QHE performance at both weak and strong system-bath coupling regimes. We have learned from Eq. (\ref{analyticalkappa}) that, Franck-Condon factor $\kappa$ is a monotonic decreasing function of temperature indicating
that higher temperature leads to strong dynamical localization. The  promotion and impediment roles that
temperature plays in energy transfer process depends on the extent of dynamical localization of the
exciton-ICT dimer system. So Fig.~\ref{uncorrelated:} can be explained as follows: at weak system-bath
couplings, the excitons are still partially dynamical delocalized, thus higher temperatures will lead to stronger
localization. Impediment effects of temperature surpass its promotion effects. When $\gamma$ becomes large
enough, the eigenstates have been fully localized. Further increasing temperature can not reduce the transfer
rate of exciton-ICT dimer system, but promote transfer process influenced by LTR1 and LTR2. Therefore, as
shown in Fig.~\ref{uncorrelated:}, when $\gamma  > 2.2$, the power gets larger with increasing
temperature.

\begin{figure}[htbp]
\centering
\includegraphics[scale=0.28]{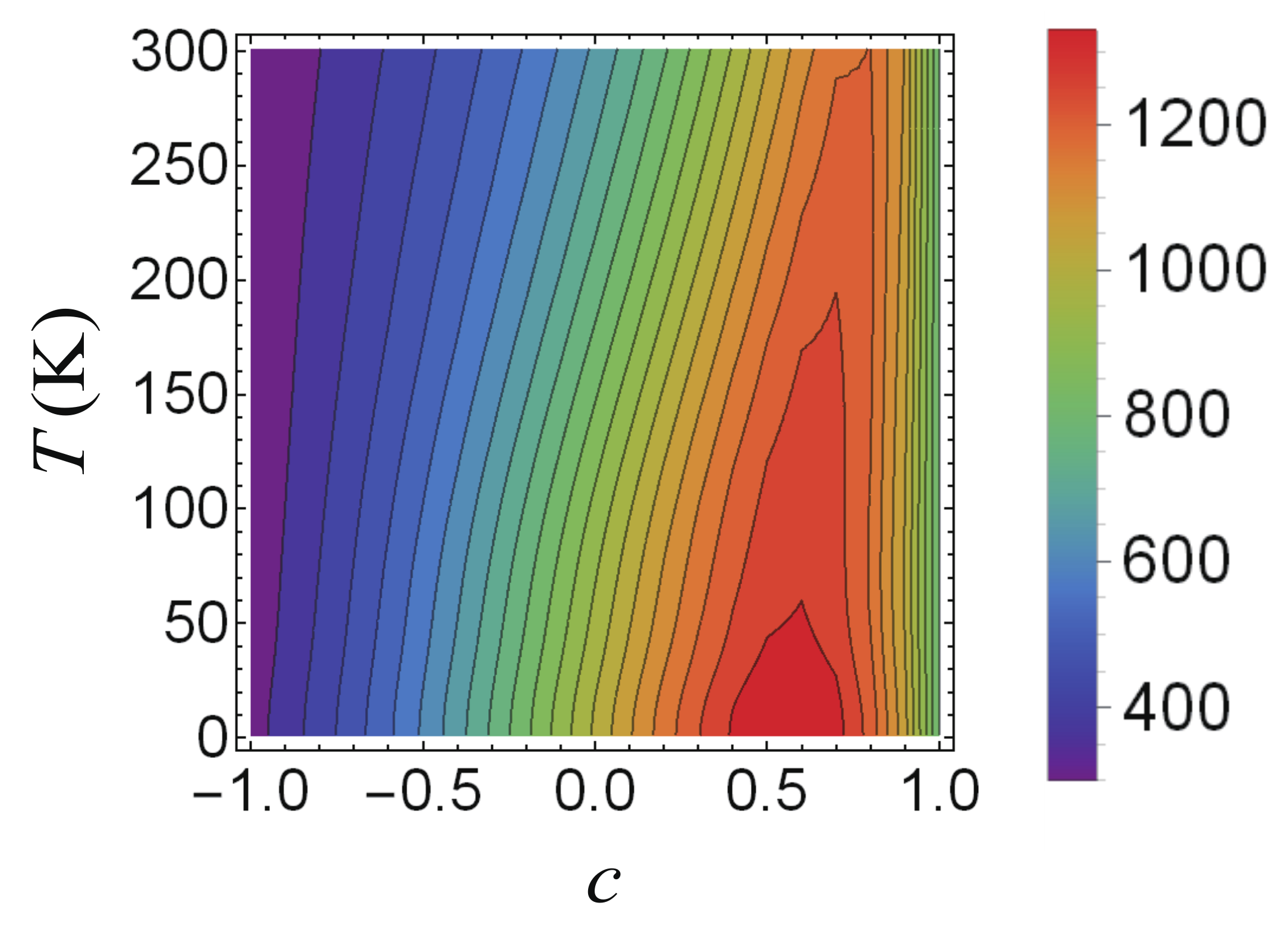}
\caption{(Color online) The power
generated by QHE as a function of temperature $T$ and cross-correlation coefficient $c $ with the system-bath coupling strength $\gamma =1$.}
\label{temperaturecrosscorrelation1:}
\end{figure}

Fig.~\ref{temperaturecrosscorrelation1:} illustrates the  power
generated by QHE as a function of temperature $T$ and cross-correlation coefficient $c$ with the system-bath coupling
$\gamma=1$. The power increases to maximal values and
then gradually decrease with $c$, in accordance with the results of Fig.~\ref{crosscorrelation:} and
Fig.~\ref{couplingstrengthcrosscorrelation:}. Similarly, temperature affects the optimal value of $c$
that yields the maximal power which also can be explained as the effect of temperature on the
balance between noise-assisted transport and dynamical localization. Meanwhile, the power
monotonically decreases with temperature. This is because the eigenstates of the exciton-ICTS dimer syatem are still partially dynamical delocalized for the given range of
parameters ($\gamma =1$ and $ - 1 \le c \le 1$), higher temperature only leads to enhanced fluctuations that gives rise to strong dynamical
localization. Although increasing $T$ promotes effective transfer rates concerned with concerned with LTR1
and LTR2, it is not not as remarkable as the impediment effects induced by strong dynamical
localization.


\subsection{Effective voltage $V$}

\begin{figure}[htbp]
\centering
\includegraphics[scale=0.17]{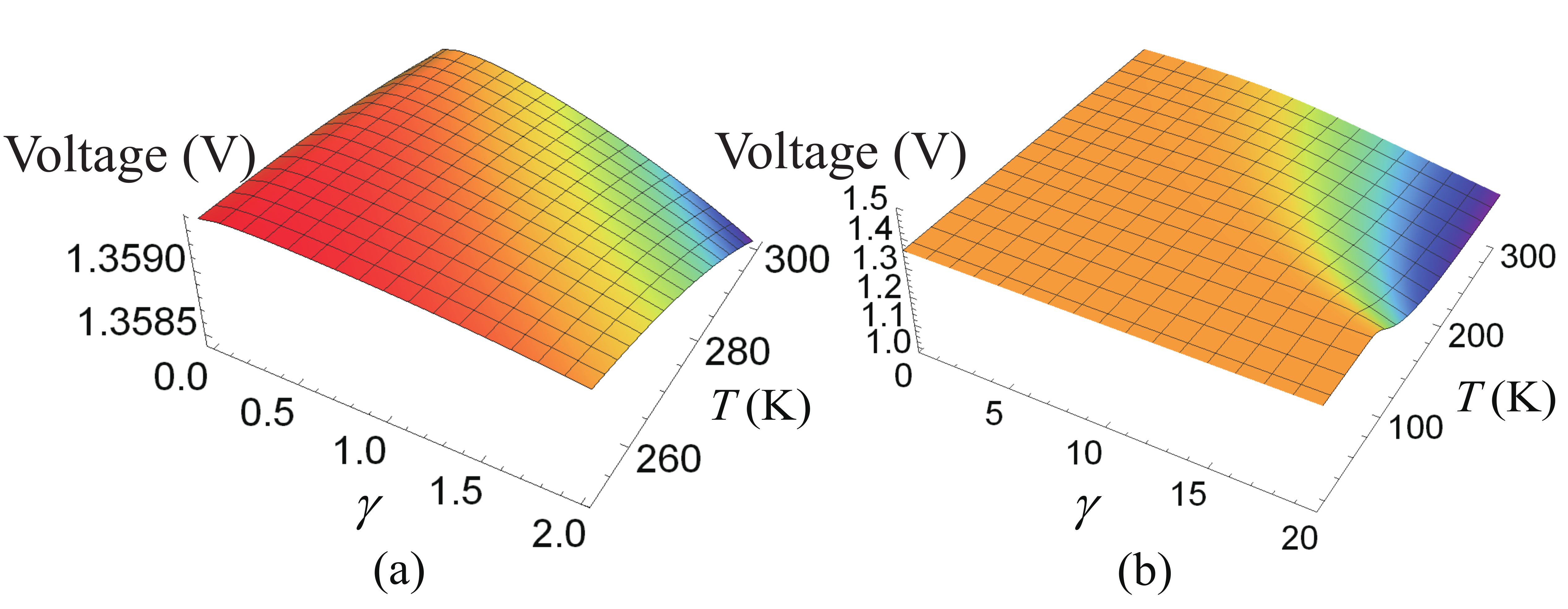}
\caption{(Color online) The effective voltage of QHE as a function of temperature $T$ and system-bath coupling strength $\gamma $ for the uncorrelated bath ($c=0$) with different parameter ranges for (a) and (b).}
\label{voltage:}
\end{figure}

Having investigated the power as a function of system-bath coupling $\gamma $, cross-correlation coefficient $c$ and temperature $T$, another question arises: how is the effective voltage $V$ influenced by these parameters? Since $c$ has been incorporated in the effective system-bath coupling $y$, effectively strengthening or weakening $\gamma $, we only focus on the impact of temperature and system-bath coupling on the voltage $V$. The results are illustrated in Fig.~\ref{voltage:}. At high temperature ($T=300$ K for example in Fig.~\ref{voltage:} (a)), the voltage increases very slightly to a maximum at first, and then decreases gradually just as the behaviors of the power with $\gamma$ and the physics is the same. But the changing rate is much slower when $\gamma $ is not very large. The effective voltage of our QHE model exhibits superior robustness with respect to the bath conditions. Thus in Fig.~\ref{couplingstrengthcrosscorrelation:}-Fig.~\ref{temperaturecrosscorrelation1:}, we only plot the power as a function of different parameters. Nevertheless, with $\gamma$ increasing further ($\gamma  > 5$), the voltage decreases almost linearly with $\gamma$ (Fig.~\ref{voltage:} (b)) indicating the population on the secondary charge separation state $\left| \alpha  \right\rangle $ decreases exponentially. In this case, energy transfer of the exciton-ICTS dimer state is completely suppressed by the extremely large fluctuations induced by phonon bath leading to sharply reduced population on the secondary charge separation state $\left| 2  \right\rangle $ and in turn ${\rho _{\alpha \alpha }} $ decreases to zero quickly. The current or power also diminishes to zero. Therefore, the population on $\left| \beta \right\rangle $ is hardly influenced by that on $\left| \alpha  \right\rangle $ but determined by the transition process subjected to LTB2 denoted by the Liouville operator ${L_{c2}}$. ${L_{c2}}$ guarantees a certain value of population on $\left| \beta  \right\rangle $, meanwhile ${\rho _{\alpha \alpha }}$ decays to zero when $\gamma$ gets extremely large. Therefore, according to the definition $eV = {E_\alpha } - {E_\beta } + {k_B}T\ln \frac{{{\rho _{\alpha \alpha }}}}{{{\rho _{\beta \beta }}}}$, it is easy to understand the variation of voltage with $\gamma$. Return to the case when $\gamma $ is not very large. From Fig.~\ref{couplingstrength:}-Fig.~\ref{temperature:}, the current $j$ (or the population ${\rho _{\alpha \alpha }}$) remains large when $\gamma < 3 $. It is ${\rho _{\alpha \alpha }}$ that mainly influences the population on $\left| \beta  \right\rangle $ by the relaxation process $\left| \alpha  \right\rangle  \to \left| \beta  \right\rangle $, rather than the transition process induced by LTB2. Thus, the ratio $\frac{{{\rho _{\alpha \alpha }}}}{{{\rho _{\beta \beta }}}}$ will not change significantly under this circumstances. In other words, $V$ is robust against the bath noise as long as  $\gamma$ is not too large. At low temperature, the tendencies are generally the same except that the voltage begins to decrease quickly at larger $\gamma$. This is because at large system-bath coupling, ${\rho _{\alpha \alpha }}$ is already very small and has an negligible impact on ${\rho _{\beta \beta }}$, but lower temperature leads to smaller population on $\left| \beta \right\rangle $ than high temperature. As a consequence, ${\rho _{\alpha \alpha }}$ should decrease to a smaller value which requires larger $\gamma$. We see that the physics underlying these results on the effective voltage $V$ also lies in the dynamical location due to bath-induced fluctuations.

\section{Conclusion}

In this paper, we apply polaron master equation to analyse the effects of system-bath couplings on charge transfer processes in PS {\rm II} RC for a wide parameter range.
Effects of bath correlations and temperature of phonon bath are also included.

Our analysis shows that temperature and system-bath coupling determine the effective electronic coupling thus influencing the equilibrium structure and dynamical localization, and in turn leading to various behaviors of QHE performance. The results reveal that system-bath coupling  can promote or impede the performance of QHE depending on the coupling strength. It is a result of a balance of noise-assisted transport and dynamical localization. The effects of bath correlation, denoted by cross-correlation coefficient $c$, is incorporated in the effective system-bath coupling $y$. Similarly, the power generated increases or decreases with
$c$ depending on the different system-bath coupling strength regimes. Temperature  affects the optimal value of $\gamma$ that yields the maximal current and power. In other words, temperature changes the balance of noise-assisted transport and dynamical localization effects. In addition, higher temperature also enhance transfer process related to LTR1 and LTR2. Whether it can promote or impede the performance of QHE depends on the extent of localization. The effective voltage $V$ changes very slowly when system-bath coupling strength is not very large, i.e., $V$ of our QHE model is robust against the bath noise. With further increasing $\gamma$,  $V$ decreases quickly which should be explained by considering the impact of dissipative phonon bath LTB2.

In summary, system-bath coupling $\gamma$, cross-correlation coefficient $c$ and temperature
$T$ conspire to affect the charge transfer process, thus showing various dynamical regimes from which two
important mechanisms dominating the transfer process in photosynthetic complexes can be extracted. The
first one is resonance energy transfer mechanism. In our model, for example, resonance energy transfer requires
the same energy of $\left| 1 \right\rangle $ and $\left| 2 \right\rangle $. This mechanism is manifested by
the noise-assisted transfer effect since the range of on-site energies of the two states are broadened due
to bath-induced energy fluctuations which leads to the overlap of $\left| 1 \right\rangle $ and $\left| 2 \right\rangle $ in energy. The second one is the dynamical localization which incorporates the effects
system-bath coupling, bath correlation and temperature. For different values of $\gamma$, $c$,
these two mechanisms interpret various dynamical regimes, and $T$ modulates the relative strength of the two mechanism. Furthermore, we attribute the two mechanisms to
one physical origin: bath-induced fluctuations. Dynamical energy fluctuations, on the one hand, gives rise to broadened on-site energies bringing about noise-induced transport effect. On the other hand, fluctuations reduce the effective electronic coupling ${\tilde J}$ leading to dynamical localization.

We hope that our work give helpful insights into
the charge transfer mechanisms in photosynthetic reaction centers and will be constructive for designing
novel nanofabricated structures for quantum transport and optimized solar cells.


\section{acknowledgments}
This work is supported by National Natural Science Foundation of
China (NSFC) under Grants No. 11175032, and No. 61475033.

\appendix
\section{The derivation of Eq.~(\ref{masterequation})}
Within the polaron frame, the thermal average of ${{\tilde H}_{sb}}$ equals to $0$, therefore the second-order perturbation theory can be applied.

To simplify further analysis, it is convenient to move into the renormalized exciton basis in which ${{\tilde H}_s}\left|  \pm  \right\rangle  = {\varepsilon _ \pm }\left|  \pm  \right\rangle$ and  transform into the interaction picture. With the Born-Markov approximation, the master equation for EET between $\left| 1 \right\rangle $ and $\left| 2 \right\rangle $ in the interaction picture is given by
\begin{eqnarray}
\begin{aligned}
\frac{{{\rm{d}}{{\rho '}^I}_s(t)}}{{{\rm{d}}t}} =  - \int_0^\infty  {{\rm{d}}s{\rm{Tr}}\{ [{{\tilde H}_{sb}}(t),[{{\tilde H}_{sb}}(t - s),{{\rho '}^I}_s(t) \otimes {{\rho '}_b}]]\} },
\label{MarkovianMasterEquation}
\end{aligned}
\end{eqnarray}
where ${\tilde H_{sb}}(t) = {\rm{ }}{e^{i({{\tilde H}_s} + {{\tilde H}_b})t}}{\tilde H_{sb}}{e^{ - i({{\tilde H}_s} + {{\tilde H}_b})t}}$.
Substitute the interaction Hamiltonian ${{\tilde H}_{sb}}(t)$ into Eq. (\ref{MarkovianMasterEquation}), and then transform back into the Schr\"{o}dinger picture and finally obtain Eq. (\ref{masterequation}):
\begin{eqnarray}
\begin{aligned}
\frac{{{\rm{d}}{{\rho '}_s}(t)}}{{{\rm{d}}t}} =&  - i\left[ {{{\tilde H}_s},{{\rho '}_s}(t)} \right] - \sum\limits_{i,j = z, \pm } {[\Gamma _{ij}^ + {\tau _i}{\tau _j}{{\rho '}_s}(t)}\\
&  + \Gamma _{ji}^ - {{\rho '}_s}(t){\tau _j}{\tau _i} - \Gamma _{ji}^ - {\tau _i}{{\rho '}_s}(t){\tau _j} - \Gamma _{ij}^ + {\tau _j}{{\rho '}_s}(t){\tau _i}].
\end{aligned}
\end{eqnarray}
Here, the Pauli operators ${\tau _i}$ are defined by Eq. (\ref{Paulioperators}) in which
\begin{eqnarray}
\begin{aligned}
 \left|  +  \right\rangle  = \cos \frac{\theta }{2}\left| 1 \right\rangle  + \sin \frac{\theta }{2}\left| 2 \right\rangle , \\
 \left|  -  \right\rangle  = \sin \frac{\theta }{2}\left| 1 \right\rangle  - \cos \frac{\theta }{2}\left| 2 \right\rangle , \\
\label{newbasis}
\end{aligned}
\end{eqnarray}
with $\tan \theta  = \frac{{2\kappa J}}{{\tilde \varepsilon }}$ and ${\tilde \varepsilon  = {{\tilde \varepsilon }_1} - {{\tilde \varepsilon }_2}}$.
The time-dependent rates $\Gamma _{ij}^ \pm $ can be calculated via the bath correlation functions
\begin{eqnarray}
\begin{aligned}
 \Gamma _{ij}^ +  = \frac{{{J^2}}}{4}\int_0^\infty  {{\rm{d}}s\left\langle {{\xi _i}(0){\xi _j}( - s)} \right\rangle } , \\
 \Gamma _{ij}^ -  = \frac{{{J^2}}}{4}\int_0^\infty  {{\rm{d}}s\left\langle {{\xi _i}( - s){\xi _j}(0)} \right\rangle } , \\
\label{rates}
\end{aligned}
\end{eqnarray}
with
\begin{eqnarray}
\begin{aligned}
 {\xi _z}(t) =& \sin \theta \left( {\tilde B(t) + {{\tilde B}^\dag }(t)} \right), \\
 {\xi _ + }(t) =& {{\tilde B}^\dag }(t)(1 - \cos \theta ){e^{i\Delta \varepsilon t}} - \tilde B(t)(1 + \cos \theta ){e^{i\Delta \varepsilon t}}, \\
 {\xi _ - }(t) =& \tilde B(t)(1 - \cos \theta ){e^{ - i\Delta \varepsilon t}} - {{\tilde B}^\dag }(t)(1 + \cos \theta ){e^{ - i\Delta \varepsilon t}}, \\
\label{xis}
\end{aligned}
\end{eqnarray}
where
\begin{eqnarray}
\begin{aligned}
\Delta \varepsilon {\rm{ = }}{\varepsilon _ + } - {\varepsilon _ + } = \sqrt {4{J^2}{\kappa ^2} + {{\left( {{{\tilde \varepsilon }_1} - {{\tilde \varepsilon }_2}} \right)}^2}},
  \label{energydifference}
\end{aligned}
\end{eqnarray}
and $\tilde B(t) = B(t) - \kappa$ with
\begin{eqnarray}
\begin{aligned}
B(t) =& {e^{i{{\tilde H}_{sb}}t}}B{e^{ - i{{\tilde H}_{sb}}t}} \\
=& \exp [\sum\nolimits_k {{{\left( {\delta {g_{vk,12}}b_k^\dag {e^{i{\omega _{vk}}t}} - \delta g_{vk,12}^ * {b_k}{e^{ - i{\omega _{vk}}t}}} \right)} \mathord{\left/
 {\vphantom {{\left( {\delta {g_{vk,12}}b_k^\dag {e^{i{\omega _{vk}}t}} - \delta g_{vk,12}^ * {b_k}{e^{ - i{\omega _{vk}}t}}} \right)} {{\omega _{vk}}}}} \right.
 \kern-\nulldelimiterspace} {{\omega _{vk}}}}} ]
  \label{bathoperatortime}
\end{aligned}
\end{eqnarray}

We need to transform Eq.~(\ref{masterequation}) back to the local frame. Since ${\sigma _z}$ commutes with the polaron transformation operator, the population term is easily to be obtained:
\begin{eqnarray}
\begin{aligned}
{\rho _{11}}(t)& = T{r_{s + b}}\left( {{\rho _{{\rm{tot}}}}\left( t \right)\left| 1 \right\rangle \left\langle 1 \right|} \right) \\
  =& T{r_{s + b}}\left( {{e^{i{\sigma _z}B/2}}{{\rho '}_{{\rm{tot}}}}\left( t \right){e^{ - i{\sigma _z}B/2}}\left| 1 \right\rangle \left\langle 1 \right|} \right) \\
  =& T{r_{s + b}}\left( {{{\rho '}_{{\rm{tot}}}}\left( t \right){e^{ - i{\sigma _z}B/2}}\left| 1 \right\rangle \left\langle 1 \right|{e^{i{\sigma _z}B/2}}} \right) \\
  =& T{r_{s }}\left( {{{\rho '}_s}\left( t \right)\left| 1 \right\rangle \left\langle 1 \right|} \right).
\label{rho11}
\end{aligned}
\end{eqnarray}
where ${{\rho '}_{{\rm{tot}}}}\left( t \right) = {e^{ - i{\sigma _z}B/2}}{\rho _{{\rm{tot}}}}\left( t \right){e^{i{\sigma _z}B/2}}$ is the polaron-transformed density matrix for the dimer system and LTB3 with ${{\rho _{{\rm{tot}}}}\left( t \right)}$ the total density matrix in the local frame and in the same way,
\begin{eqnarray}
\begin{aligned}
{\rho _{22}}(t)= T{r_{s + b}}\left( {{{\rho '}_s}\left( t \right)\left| 2 \right\rangle \left\langle 2 \right|} \right).
\label{rho22}
\end{aligned}
\end{eqnarray}
Nevertheless, ${\sigma _x}$ (${\sigma _y}$) does not commute with the polaron transformation operator. Consequently, the exact coherence terms can not be obtained. We can use the Born approximation ${{\rho '}_{{\rm{tot}}}}\left( t \right) \approx {{\rho '}_s}\left( t \right) \otimes {{\rho '}_b}$ to solve this problem. The Born approximation has already been
used in deriving Eq.~(\ref{masterequation}, assuming that the coupling between the system and bath is weak.
As in the polaron transformation theory, the system-bath interaction term is reduced and can be treated as
a perturbation, it is reasonable to be approximately factorize the density matrix in the polaron frame.
Therefore, we obtain the coherence term of the reduced system density matrix in the local frame as
\begin{eqnarray}
\begin{aligned}
{\rho _{12}}(t) &= T{r_{s + b}}\left( {{\rho _{{\rm{tot}}}}\left( t \right)\left| 2 \right\rangle \left\langle 1 \right|} \right) \\
  =& T{r_{s + b}}\left( {{e^{i{\sigma _z}B/2}}{{\rho '}_{{\rm{tot}}}}\left( t \right){e^{ - i{\sigma _z}B/2}}\left| 2 \right\rangle \left\langle 1 \right|} \right) \\
  =& T{r_{s + b}}\left( {{{\rho '}_{{\rm{tot}}}}\left( t \right){e^{ - i{\sigma _z}B/2}}\left| 2 \right\rangle \left\langle 1 \right|{e^{i{\sigma _z}B/2}}} \right) \\
  \approx &T{r_{s + b}}\left( {{{\rho '}_s}\left( t \right) \otimes {{\rho '}_b}{e^{ - iB}}\left| 2 \right\rangle \left\langle 1 \right|} \right) \\
  =& \kappa T{r_s}\left( {{{\rho '}_s}\left( t \right)\left| 2 \right\rangle \left\langle 1 \right|} \right).
\label{rho12}
\end{aligned}
\end{eqnarray}

\end{document}